\documentclass[onecolumn,compsoc,journal]{IEEEtran}
  \RequirePackage[compress]{cite}
  \input{a00-preamble.tex}

  \newcommand{\citep}{\cite}
  \newcommand{\citet}{\cite}

  \usepackage{color,array,verbatim,algpseudocode,bm}
  \definecolor{menucolor}{rgb}{0.1,0.52,0.47}
  \definecolor{urlcolor}{rgb}{0.85,0.37,0.01}
\definecolor{runcolor}{rgb}{0.46,0.44,0.701}
\definecolor{filecolor}{rgb}{0.2,0.5,0.01}
\definecolor{linkcolor}{rgb}{0.12,0.47,0.70}
\definecolor{citecolor}{rgb}{0.55,0.36,0.01}
\definecolor{anchorcolor}{rgb}{0.4,0.4,0.4}
\usepackage[colorlinks=true, urlcolor=urlcolor, linkcolor=linkcolor,citecolor=citecolor,filecolor=filecolor, anchorcolor=anchorcolor,menucolor=menucolor]{hyperref}
\usepackage{nameref}
\usepackage{zref-xr}
\zxrsetup{
  tozreflabel=false,
  toltxlabel=true,
}

\usepackage[nodisplayskipstretch]{setspace}

\makeatletter
\newcommand\VeryLarge{\@setfontsize\Huge{16}{16}}
\newcommand\smaller{\@setfontsize\small{4}{4}}
\makeatother   

\IEEEaftertitletext{\vspace{-\baselineskip}} 



\normalsize

\begin{document}
\title{
Optimized $k$-means color quantization of digital images in machine-based and human perception-based colorspaces}
\author{
  Ranjan Maitra\thanks{R. Maitra is with the Department of Statistics, Iowa State University, Ames, Iowa, USA.}
}
\maketitle
\begin{abstract}
Color quantization represents an image using a fraction of its original number of colors while only minimally losing its visual quality. The $k$-means algorithm is commonly used in this context, but has mostly been applied in the machine-based RGB colorspace composed of the three primary colors. However, some recent studies have indicated its improved performance  in human perception-based colorspaces. We investigated the performance of $k$-means color quantization at four quantization levels in the RGB, CIE-XYZ, and CIE-LUV/CIE-HCL colorspaces, on 148 varied digital images spanning a wide range of scenes, subjects and settings. The Visual Information Fidelity (VIF) measure numerically assessed the quality of the quantized images, and showed that in about half of the cases, $k$-means color quantization is best in the RGB space, while at other times, and especially for higher quantization levels ($k$), the  CIE-XYZ colorspace is where it usually does better. There are also some cases, especially at lower $k$, where the best performance is obtained in the CIE-LUV colorspace. Further  analysis of the performances in terms of the distributions of the hue, chromaticity and luminance in an image presents a nuanced perspective and characterization of the images for which each colorspace is better for  $k$-means color quantization.

\end{abstract}
\begin{IEEEkeywords}
brightness; circular statistics; multivariate random forest; multivariate regression tree; skewness; white point
\end{IEEEkeywords}



\section{Introduction}
\label{S:1}
True color images are often represented with high fidelity in rasterized form by lossless formats such as the Portable Network Graphics (PNG) or the Tagged Image File Format~\citep{murrayandvanryper96} (TIFF). The rasterized image is
 a grid of pixels, or picture elements~\citep{hinchcliffe2003use} that is typically read in from left to right and top to bottom. 
 The file format uses a 24-bit pixmap (or a spatially mapped array of
 3 unsigned character bytes) to represent color at each
 pixel~\citep{acharya2002integrated}. Color in each pixel can be
 represented in terms of its primary components, that is, red, 
green and blue, or RGB in abbreviated form~\citep{goldberg1991colour}. The model underlying the RGB colorspace is additive in that the color at each pixel is a composition of different intensities of the RGB colors. This framework underlies how color is specified at each pixel of a digital image. Thus, zero intensity for the three colors results in a black pixel while the three colors, each at maximum intensity, yield a white pixel. The intensities of red, green, and blue each have a total of $2^8$ colors, ranging from 0 to 255. 
Thus,  any digital image can be displayed using the RGB 3D colorspace model at each pixel~\citep{joblove1978color}.  However,
representing color is  subjective and indeed, matching 
the human brain's perception of color is a complex
task~\citep{ford1998colour}. The International
Commission of Illumination (CIE)  therefore developed colorspaces
({\it e.g.}, CIE-XYZ or XYZ, CIE-LUV or LUV and CIE-HCL or HCL)
beyond the RGB space to better mimic in a digital image the way in which the human brain perceives color. 

Digital images potentially contain millions of colors that can
nowadays be fairly easily displayed on 24-bit hardware, however it may still
be more efficient and desirable to display, store or transmit them with fewer colors, while losing as little visual information as possible~\citep{yang1998color,celebi2011improving,maitra2012bootstrapping}. 
This reduced representation of an image is called color
  quantization. In general, and as discussed in a recent review~\citep{celebi23} of the state of the art in color quantization of images, there are two broad sets of approaches: the image-independent methods that use color
from a reduced but fixed palette, and adaptive methods that decide on a
palette based on the distribution of colors in an image~\citep{xiang1997color}.
There are many image-dependent methods\citep{celebi09,celebiandschaefer10,schaeferetal11,ozdemirandakarun02,schaeferandzhou09,bingetal04,dekker94,papamarkosetal02,changetal05,boundsetal24}:
of particular interest for us in this paper 
is the  $k$-means algorithm~\citep{hartigan1979algorithm,lloyd82} for
color quantization~\citep{verevkaandbuchanan95,celebi2011improving}
  which has been shown to exhibit very satisfactory
  performance~\citep{scheunders97}.

  The use of $k$-means  and other color quantization
  approaches has typically been in the RGB colorspace. However, it is
  worth investigating if $k$-means color quantization can be improved when applied  in other colorspaces.   In this paper,
  therefore, we investigate its performance in the RGB as well as in the alternative
  XYZ, LUV and HCL colorspaces. Though recent work~\citep{celebietal23} has provided a high-quality image dataset for color quantization research, our  investigation is on 148 different images that are of different and varied scenes, settings and lighting conditions, as characterized by the distribution of the hue, chromaticity and luminance of the image pixels.   Performance in color quantization and   image compression is usually measured in terms of the
  Peak-Signal-to-Noise-Ratio~(PSNR)~\citep{delgadoandcelebi24}, but this is not a fair comparison given that the $k$-means optimization function is a simple calculation on the PSNR, so we instead evaluate performance in terms of the Visual Information Fidelity~(VIF)~\citep{sheikhetal05,sheikhetal06} that is a popular and agnostic measure that simply compares the original image with its quantized version  using natural scene statistics. We see that on the whole, $k$-means color quantization is best in about half of the cases in the RGB colorspace, while for the rest, XYZ is the best, especially at higher $k$.   Further analysis of the performance in terms of the distributional characteristics of the hue, chromaticity and luminance in an image provides a refined understanding of when each colorspace is a better performer with regard to $k$-means color quantization. 

The main paper has three more sections. Section~\ref{sec:methods} defines and describes the different colorspaces, the $k$-means algorithm, the VIF, and introduces and characterizes the images in terms of the distribution of the hue, chromaticity and luminance in the image pixels. Section~\ref{sec:results} details and analyzes the results of our investigation. We  conclude with some discussion. Because of the diversity in our publicly available images, and detailed distributional characterization of each image's hue, chromaticity and luminance, our paper includes, in an appendix, a short description of each of the 148 images used  in our investigations.
\section{Materials and Methods}
\label{sec:methods}
\subsection{Background and preliminaries}
\subsubsection{Colorspaces and transformations}

The human brain perceives color through the interaction of a stimulus that is transported from the receptors of the eye to the brain~\citep{sharma2002digital}. Color perception therefore begins in the back of the eye, where there are three cone and two rod 
cells. For humans with normal vision, the three cone cells 
perceive visible color by the frequencies or intensities of the
wavelengths produced by the three primary colors while the rod
cells are more active under diminished light conditions. The CIE  concluded in
1931 that the average human visual system is only able to
detect light between approximately 360 nm and 830 nm in
wavelength~\citep{schanda07}. The spectral sensitivity, or the relative efficiency of
detecting light, peaks at
wavelengths of approximately 430 nm for blue light, 535 nm for green light, and
570 nm for red light. Therefore, while color can be represented in 
3D space,  there are wavelengths of electromagnetic radiation outside of the spectral
sensitivity of the human visual system. For instance, ultraviolet light has wavelengths of between 10 nm and 
380 nm that are well outside the perceptive capacity of human vision. 
The RGB colorspace does not cover the whole gamut of human color perception, despite being the most commonly used colorspace in computer software. Other  drawbacks of the RGB space include the fact that it does not describe color in  terms of its hue, brightness, and colorfulness, which is how humans perceive color. It is also a non-linear colorspace that sometimes creates optical illusions. So we introduce other colorspaces in the digital imaging domain that lie inside the visible spectrum and more closely match human perception.

The CIE's first attempt in 1931 to represent color in terms of hue, brightness and colorfulness, led to the development of the CIE-XYZ colorspace that,  unlike RGB, is more in tune with human perception.  However, before conversion from RGB to XYZ space, the type of brightness and the chromaticity to be used~\citep{lindbloom1989accurate} needs to be specified. In particular, how a human perceives color is defined by
\begin{itemize}
\item \textit{luminance}, which is an indicator of the brightness or light used, and 
\item \textit{chromaticity}, the specification of the color independent of the luminance or light used. Chromaticity  is usually defined by its hue and colorfulness. 
\end{itemize}

In specifying human perception-based colorspaces, we also  need to account for the fact that humans perceive color differently according to the source of light. Therefore, many colorspaces need the specification of a white point in order to
transform color from RGB. For instance, the white point known as
D65, which is also what we use in this paper, estimates 
the color (white) produced by sunlight  at mid-day. (An alternative illuminant
reference white point that models how
humans detect white in fluorescent light is D50.)

The conversion of RGB values to its linear form (standard RGB, or sRGB) is through a non-linear transfer function which is the combination of a linear function at low brightness values and a displaced power law for the rest of the range. This function is the same for all three channels and is specified as
\begin{equation}
  x'={\begin{cases}\frac x{12.92},&x\leq 0.04045\\\left({\frac {\displaystyle x+0.055}{\displaystyle 1.055}}\right)^{2.4},&x>0.04045\end{cases}}
  \end{equation}
for each  $x\in\{R,G,B\}$ with $R',G',B'$ being the values in the sRGB space, and where it is assumed that the RGB values are each scaled to be in [0,1].

The sRGB tristimulus values are then converted~\citep{hardeberg01} to XYZ values through a linear transformation 
specified by a $3{\times}3$ matrix that  is chosen based on the luminance reference angle and reference white point. 
The angle used for our conversion formulae is $2^\circ$ and the reference white point is  D65. With this white point (D65) and $2^\circ$ luminance perception, we arrive at the formula for converting sRGB to XYZ:
\begin{equation}
  \left(\begin{array}{c}
X\\
Y\\
Z
  \end{array}\right)=\left[\begin{array}{ccc}
0.4124564 & 0.3575761 & 0.1804375\\
0.2126729 & 0.7151522 & 0.0721750\\
0.0193339 & 0.1191920 & 0.9503041
  \end{array}\right]\left(\begin{array}{c}
R'\\
G'\\
B'
  \end{array}\right)
.
  \label{eq:RGBtoXYZ}
\end{equation}

From \eqref{eq:RGBtoXYZ}, it is easy to see that each component of the XYZ colorspace is a linear combination of different RGB intensities. Further, the $Y$ component approximately represents the brightness in a color. The contrast between colors in an image works differently with different brightness, applied to each pixel of the image. Those pixels with higher $Y$ values are brighter while colors with approximately equal $Y$ intensity values are perceived by the human eye with the same amount of brightness~\citep{mendoza2006calibrated}. In other words, the colorspaces that separate  brightness and chromaticity perform differently than in the RGB colorspace when quantized. 
Further, for the reference white point D65, the reference XYZ values are given by
\begin{equation}
  (X_r,Y_r, Z_r) = (95.5, 100, 108.9)
  \label{eq:XYZD65}
\end{equation}

The CIE also created a 2D  $xy$ chromaticity space to describe color, where  chromaticity is defined, independently of the brightness, as a combination of the hue and colorfulness~\citep{kerr2010cie}. This separation is done once the RGB triplets are converted to XYZ triplets. To map each XYZ triplet to its  2D $xy$ chromaticity pair, we  specify the coordinates of the primary red, green, and blue colors on the chromaticity diagram~\citep{heredia1998chromatic}. From the XYZ triplets, the $xy$ chromaticity diagram is defined by lower case $xyz$ as:
\begin{equation}
  \begin{split}
  x=\frac{X}{X+Y+Z},\qquad
  y=\frac{Y}{X+Y+Z},\qquad
  z=\frac{Z}{X+Y+Z} \\
    \end{split}
  \label{eq:XYZtoxyz}
\end{equation}
From \eqref{eq:XYZtoxyz}, it is clear that the third variable is redundant since $x+y+z=1$. Thus, the one-to-one mapping of the 3D colorspace vectors to the 2D  chromaticity is based on the intensity and the wavelength of color displayed at an image pixel~\citep{schanda07}.

The CIE-XYZ space moved the representation of color to a space more aligned with human perception but continued to have drawbacks. Notably, like RGB, the XYZ space is nonlinear~\citep{celebi2010fast}. Further, it maps poorly to how humans perceive color  because the 3D values are sometimes confounded. So, in 1976, the CIE created the 
CIE-LUV space in an attempt to emulate how humans perceive brightness linearly. They sampled people with normal human vision and determined how people perceive brightness, and ended with a linear approximation that is the luminance of the colorspace. The relation for luminance $L$ is in terms of $Y$, and given by
\begin{equation}
  L=\begin{cases}{\bigl (}{\tfrac {29}{3}}{\bigr )}^{3}\frac{Y}{Y_{r}},&\frac{Y}{Y_{r}}\leq {\bigl (}{\tfrac {6}{29}}{\bigr )}^{3},\\116{\sqrt[{3}]{\frac{Y}{Y_{r}}}}-16,&\frac{Y}{Y_{r}}>{\bigl (}{\tfrac {6}{29}}{\bigr )}^{3}.\end{cases}
    \label{eq:luminance}
\end{equation}
Luminance is, however, just one aspect of how humans perceive color so we also need to specify chromaticity. In other words, we need to define how the human brain is able to differentiate between the different colors or different brightnesses. The 2D vector $(u, v)$ in the CIE-LUV colorspace represents chromaticity~\citep{schanda07} as
\begin{equation}
u=13L (u^{\prime }-u_{r}^{\prime }),\qquad v=13L (v^{\prime }-v_{r}^{\prime }),
  \label{eq:uv}
\end{equation}
where $u^\prime$ and $v^\prime$ are given~\citep{poynton03} by the equations:
\begin{equation}
  \begin{alignedat}{3}u^{\prime }&={\frac {4X}{X+15Y+3Z}}&&={\frac {4x}{-2x+12y+3}},\\v^{\prime }&={\frac {9Y}{X+15Y+3Z}}&&={\frac {9y}{-2x+12y+3}},\end{alignedat}
\end{equation}
  and $u_r^\prime$ and $v_r^\prime$ are from a reference point for white, here taken to be D65 and the $2^\circ$ field view, and for which \eqref{eq:XYZD65} yields $(u_r^\prime,v_r^\prime)=(0.19873, 0.46821)$. With this reference point, \eqref{eq:uv} reduces to 
\begin{equation*}
u=13L(u^\prime-0.19873), \qquad v=13L(v^\prime-0.46821).
\end{equation*}
For the white point and angle considered in this paper, the components $(u,v)$ take values in (-100,100). CIE-LUV is a perceptually uniform colorspace, but it does not map the colors correctly when either $u$ or $v$ is 0 because the result then is a black pixel. A way around this is to represent these coordinates in terms of their polar coordinates, providing the CIE-HCL colorspace that keeps $L$ as in the CIE-LUV specification, but defines chromaticity and hue in terms of 
\[
C=\sqrt{u^{2}+v^{2}}, \qquad H=tan^{-1}\left(\frac vu\right),
\]
The HCL colorspace~\citep{zeileisetal09} gets around many of the deficiencies of the other colorspaces, and is considered to be a good representation of how people with normal vision perceive color. 

We conclude our discussion on colorspaces by noting that there exist other spaces such as the CIE-LAB~\citep{schanda07}, which is considered better for displaying reflective surface colors illuminated by a standard illuminant like D65. However, since our focus here is on digital 
images, we use  CIE-LUV (and HCL) that is considered to be better for emissive, self-illuminated colors~\citep{robertson90}. The latter also has stable saturation performance. Further, while both CIE-LAB and CIE-LUV use luminance L for perceptual lightness that is somewhat perceptually uniform, both of them are not particularly uniform in terms of hue and chroma, with CIE-LAB being markedly worse than CIE-LUV.  The former also suffers from unstable hue that changes due to luminance or chroma, and has significant inaccuracies in the blue area. CIE-LAB colors are also more unevenly distributed than CIE-LUV.
It is for these reasons that we do not consider the CIE-LAB space in our investigation into the colorspaces for $k$-means color quantization of digital images. 

\subsubsection{$k$-means color quantization}
Color quantization represents digital images by means of a few colors without appreciably degrading their visual quality~\citep{pappas97,xiang1997color,sirisathitkul2004color,celebi09,celebiandschaefer10,maitra2012bootstrapping,berlinskiandmaitra25}. Strictly speaking, any digital display of an image, for example in TIFF format, already quantizes an image to an extent~\citep{heckbert1982color}, but here, the discussion is on whether we can map the original digital image pixels to some more reduced representation. 
There are two stages to color quantization, that is, choosing a palette of colors and then mapping the pixels from the original  to its quantized image~\citep{hwang2002fast}. 
Indeed, there exist many methods for color quantization~(see~\citet{celebi23} for a fairly comprehensive and up-to-date review), but the focus in this paper is on the commonly-used $k$-means color quantization methods that essentially segments or partitions the trivariate pixel values in RGB space using the $k$-means clustering algorithm.
This algorithm for color quantization of a digital image is an unsupervised aapproach that, given $k$, groups the colors in an image based on the Euclidean distance metric. There are several ways to decide on $k$ in the context of $k$-means~\citep{vendraminetal10}. Specific to color quantization, \citet{maitra2012bootstrapping} provided a data-driven hypothesis testing-based method that identified both the minimal and sufficient number of colors for adequate representation of a digital image using $k$-means color quantization. However, in this paper, we assume that we have decided on a pre-specified number of colors.

The $k$-means algorithm~\citep{macqueen67,lloyd82,hartigan1979algorithm}, in the context of color quantization of images, partitions the original color values $\mX=[x_{1}, x_{2}, ..., x_{N}]$ into $k$ groups $[C_{1}, C_{2}, ..., C_{k}]$ while mimimizing 
the within cluster  sum of squares (WCSS) 
\[
  \mbox{WCSS}={\sum_{i=1}^{N}\sum_{j=1}^k||x_{i}-\mu(C_{j})||^{2}}\mathbb{I}[x_i\in C_j] ,
\]
where $\mu(C_{j})$ is the mean of the group $C_j$, and $\mathbb{I}[\cdot]$ is the indicator function.  The efficient Hartigan-Wong~\citep{hartigan1979algorithm,telgarskyandvattani10,slonimetal13}  implementation that reduces unnecessary computations is an Optimal Transfer Quick Transfer (OTQT) algorithm~\citep{dormanandmaitra22,berlinskiandmaitra25} and what we use in our investigations. Proper initialization is an important determinant of algorithm performance ~\citep{maitra09} and we use the $k$-means$++$ approach~\citep{arthur2007k} along with our OTQT-based $k$-means algorithm. 

Color quantization using $k$-means has generally been performed in the RGB space, however, it is worth investigating if the human perception-based colorspaces may provide better quantization, that is less degradation of image quality. While there has been some interest~\citep{berlinskiandmaitra25,pappas97,yoonandkweon04} in performing color quantization in human perception-based colorspaces, the literature is sparse on a comparative assessment of $k$-means or other color quantization in different computer-based or human perception-based colorspaces. It is this lacuna that we seek to remove by evaluating performance of $k$-means color quantization in the RGB, XYZ or LUV colorspaces. We note that $k$-means is not directly usable on HCL values because $H$ is represented as an angle, so we need to use a transformation for $H$ and $C$ to Euclidean space, that essentially is transforming HCL to its LUV counterpart. For angular data, the $k$-mean directions algorithm~\citep{maitraandramler10} is used as an analogue of the $k$-means algorithm for directional data, however, there is no obvious way to combine this algorithm  with $k$-means for chromaticity and luminance that take values in the Cartesian space. Therefore, we use transformations between the HCL and the LUV spaces. Fortunately for us, this is not a problem because  the $k$-means optimization function can be viewed as a Classification-Expectation Maximization solution~\citep{celeuxandgovaert92} in a Gaussian mixture model with all components having a common spherical dispersion matrix and uniform mixing proportions, and further,
maximum likelihood-based parameter estimation in each $k$-means group is unaffected by the transformation from LUV to HCL space. To see this, suppose that for each $k$-means group in the LUV space, the pixel values have density given by $f(L, U, V; \bvartheta)$ where $\btheta$ is the  (mean) parameter for the pixels assigned to that group. Then, because the transformation from LUV to HCL is $U=C\cos H$ and $V=C\sin H$ while $L$ remains unchanged, the Jacobian of the transformation is  $C$ and so the density of $(H,C,L)$ is simply $Cf(L,C\cos H,C\sin H;\bvartheta)$. Then, the estimate of $\bvartheta$ remains unchanged within each $k$-means component, and provides the theoretical underpinning for performing $k$-means in the LUV space and using that for similar optimization in the HCL-space. Therefore, we only use $k$-means in the LUV space in place of HCL for color quantization. Consequently, we investigate the performance of $k$-means color quantization on the three colorspaces. Finally, unlike in the RGB space, the three  components of XYZ or LUV do not have the same range. Our investigations therefore scale each of the three components in a colorspace to each lie inside [0,1] before applying the $k$-means algorithm on these scaled trivariate values.
\subsection{Digital images testbed}
%
Our analysis used 148 small to very large digital images, encompassing a host of subjects, lighting conditions and situations. Nine of these images are described in Section~\ref{sec:showcase} while the rest are described in the Appendix. 
All images are publicly available, whether at the link specified in the descriptions, or otherwise at \url{commons.wikimedia.org}. Our 148 images in Section~\ref{sec:showcase} and  Appendix~\ref{app:addlimages} span the gamut of sizes, scenes, lighting conditions and the goal of our study is to provide us with some understanding of the image characteristics governing the performance of $k$-means color quantization in the different colorspaces.

\subsubsection{Characterizing an image in terms of the distribution of its hue, chromaticity and luminance}

  As can be seen in Appendix~\ref{app:addlimages}, and together with the nine showcase images, the 148 images span the gamut of sizes, scenes, lighting conditions and can provide us with some understanding on the conditions under which the different colorspaces are related to better performance of $k$-means color quantization.
\begin{figure}
\includegraphics[width=\textwidth]{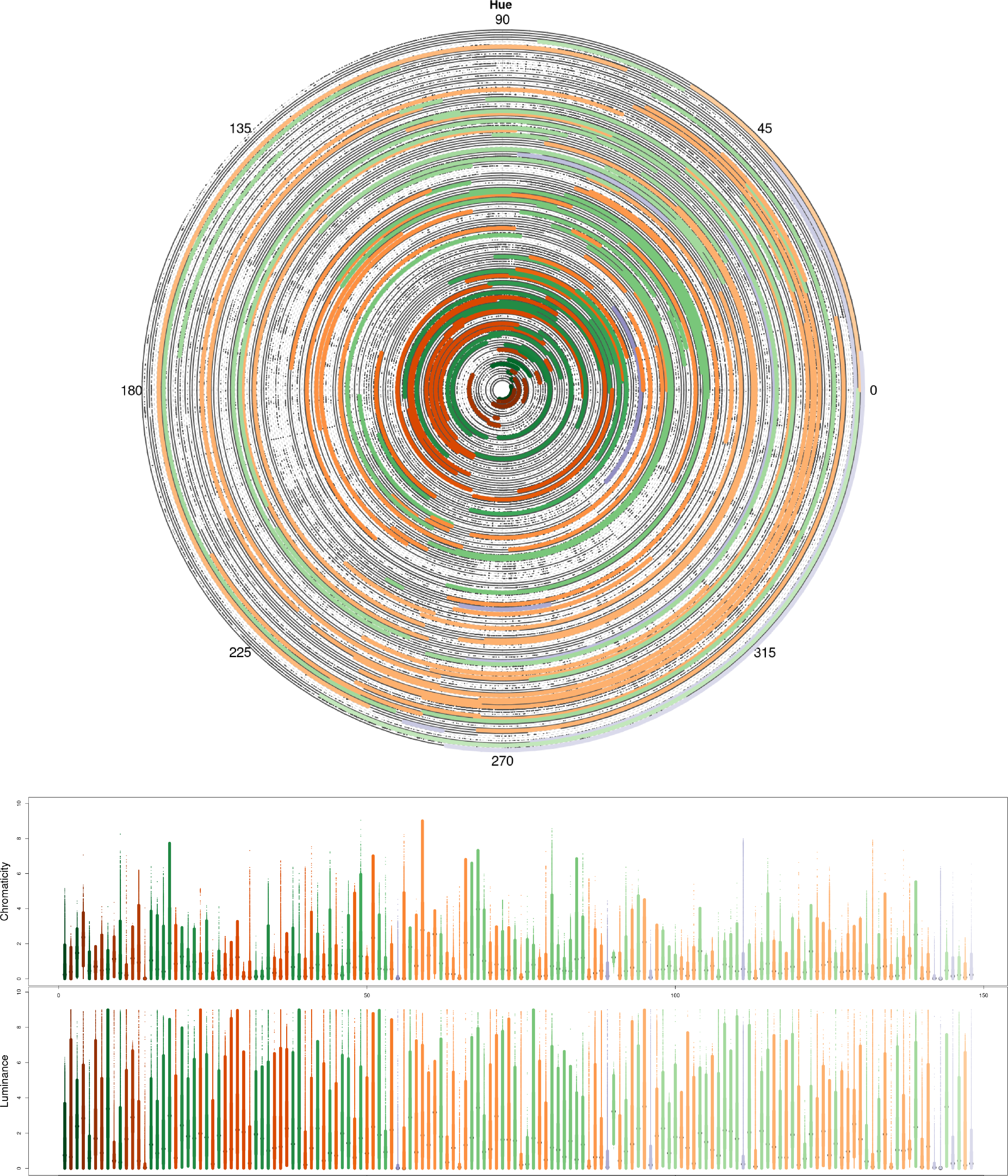}
\caption{Distributions of the hue, chromaticity and luminance in  each image used in our study. The distributions are displayed by means of grouped circular boxplots (for hue) and linear boxplots for chromaticity and luminance. The distributions, and the brightness of the colors in the display are ordered from inside out, and left-to-right in order of number of image pixels. The orange, green and purple palettes represent the distributions for the images where 64-means color quantization did best using RGB, XYZ and LUV colorspaces, respectively.}
  \label{fig:hcldist}
\end{figure}
The variety in the images considered is displayed in Figure~\ref{fig:hcldist} in terms of the distributions of the hue, chromaticity and luminance at each pixel. 
We display the distribution of hue (an angular quantity), chromaticity and luminance (which are linear quantities) through circular~\citep{buttarazzi18,berlinskietal25} and linear~\citep{tukey77} quartile plots, which are analogues of boxplots, originally defined in the linear case by  \citet{tufte98} and extended to the circular case by \citet{berlinskietal25} for displaying a large number of distributions. 

\subsubsection{Summarizing the distribution of the hue, chromaticity and luminance in an image}
We further characterize each distribution of hue, chromaticity and luminance for ready reference in terms of the four statistical measures of mean, standard deviation, skewness and kurtosis~\citep{kendallandstuart69}. While not exhaustive, these four measures provide a fairly comprehensive description of a distribution so we discuss and use them in our analysis. 
Also these measures are differently described for linear observations (such as chromaticity or luminance) and for directional or angular observations (such as hue) so we define them separately for these two scenarios. 

\paragraph{Linear statistics}
The mean of a distribution is a measure of location of its center, and is computed from a sample $X_1,X_2,\ldots,X_n$  by the formula $\hat\mu = \frac1 n\sum_{i=1}^nX_i$. The standard deviation, or SD, of the distribution is a measure of its spread (or deviation) from the mean and is calculated for a sample from the formula $\hat\sigma= \sqrt{\frac1{n-1}\sum_{i=1}^n (X_i -\bar X)^2}$. The skewness of a distribution measures its asymmetry about the mean, and is calculated for a sample from the Fisher-Pearson formula of standardized third moment~\citep{doaneandseward11} expressed as $\hat\zeta = \frac{m_3}{m_2^{3/2}}$ where $m_l$ is the $l$th central moment of the data and calculated using the formula $m_l = \frac1{n}\sum_{i=1}^n (X_i -\bar X)^l$. In general, $\hat\zeta$ can be negative (for a negatively-skewed distribution where the left tail is longer relative to the right), positive for a positively-skewed distribution, with longer right tail than the left, or zero (which is commonly attained, for example, for a symmetric unimodal distribution). Our fourth measure of summary of a linear distribution is its kurtosis or measure of the extremity of its tails~\citep{westfall14}. 
The Pearson formula~\citep{pearson1905} for sample kurtosis is provided by $\hat\kappa = \frac{m_4}{m_2^2}$ and is an average of the standardized sample raised to the fourth power. In the formula, standardized values less than unity correspond to data within 1 SD of the mean and contribute minimally to $\hat\kappa$ while the meaningful contributions to the measure are from data values farther away from the mean. Thus, we  see that outliers drive the value of this measure and the formula itself provides an easy illustration of why the definition of the measure provides an indication of the extremity of the outliers in the dataset. 

\paragraph{Circular statistics} Circular (or angular) data provide a complication in that usual measures and analysis are not always directly applicable. This is because the complication introduced by the periodicity (wraparound) of the angle  needs to be addressed. We illustrate this point through a simple example where we have only two angular observations ($1^\circ$ and $359^\circ$). Then the usual (linear) mean formula, ignoring the periodicity and wraparound effect,  would yield a mean of $180^\circ$ which is actually entirely opposite the true (and intuitive) mean direction value of $0^\circ$. This simple illustration shows the need for  care in summarizing and analyzing directional data and is provided through a separate set of measures defined for directions and angles~\citep{mardiaandjupp99} that we  describe now.
\newline
\indent\indent\indent\indent\underline{\bf Mean direction}: The mean direction, also called the angular or circular mean, is a measure of location for directional observations. For a sample $\theta_1,\theta_2,\ldots,\theta_n$ of angular observations, the mean direction is defined~\citep{mardiaandjupp99} to be the $\hat\mu^\circ\equiv\bar\theta\in[0^\circ,360^\circ)$ satisfying the  two equations 
\begin{equation*}
  \bar C =\frac1n \sum_{j=1}^n\cos\theta_j = \bar R\cos\bar\theta , \qquad  \bar S =  \frac1n\sum_{j=1}^n\sin\theta_j =\bar R\sin\bar\theta, 
\end{equation*}
with 
$\bar R = \sqrt{\bar C^2 + \bar S^2}$ defined as the mean resultant length of the sample. 
The sample mean direction is equivariant under rotation, which means that two practitioners using different coordinate systems will arrive at the same answer for the mean direction, even though they may use different descriptions to define describe its actual position. 
This equivariance property under rotation of the mean direction is analogous to the property of equivariance under translation of the sample mean for linear data. Finally, we point out that the mean direction is the only statistic  used in our analyses that is an angular quantity, and constrained to be in $[0^\circ,360^\circ)$: the other three statistics for circular data used in this paper do not have this constraint. 
\newline
\indent\indent\indent\indent\underline{\bf Circular standard deviation}:
The mean resultant length $\bar R$ is a measure of concentration of a dataset of angular observations. If the observations $\theta_1,\theta_2,\ldots,\theta_n$ are all very close to each other, then $\bar R$ is close to unity, while  it is close to zero if these observations are very widely dispersed. However, this diagnosis is not definitive: \citet{mardiaandjupp99} point out through an example with $2n$ observations 
 $\{\theta_1,\theta_2,\ldots,\theta_n, \theta_1+180^\circ,\theta_2+180^\circ,\ldots,\theta_n+180^\circ\},  $ that the sample has $\bar R=0$, regardless of the actual values of $\theta_1,\theta_2,\ldots,\theta_n$, and regardless of how spread or close these observations are relative to each other. This illustrates why having 
$\bar R \approx 0$ does not necessarily imply that the directional observations are spread almost evenly around a circle. The circular standard deviation (SD) is defined as $\hat \sigma^\circ = \sqrt{-2\log\bar R}$~\citep{mardiaandjupp99}.
\newline
\indent\indent\indent\indent\underline{\bf Circular skewness}:
The $p$th trigonometric moment is used to define measures of circular skewness and kurtosis. For a sample $\theta_1,\theta_2,\ldots,\theta_n$, the $p$th trigonometric moment is defined in terms of the sample mean direction $\hat\mu^\circ_p$ and mean resultant length $\bar R_p$ of $p\theta_1,p\theta_2,\ldots,p\theta_n$. Specifically, $\hat\mu^\circ_p\equiv\bar\theta_p\in[0^\circ,360^\circ)$ and $\bar R\in[0,1]$ satisfy 
\begin{equation*}
  \bar C_p =\frac1n \sum_{j=1}^n\cos p \theta_j = \bar R_p\cos\bar\theta_p , \qquad  \bar S_p =  \frac1n\sum_{j=1}^n\sin p\theta_j =\bar R_p\sin\bar\theta_p.
\end{equation*}
The circular skewness is then defined as
\begin{equation*}
\hat\zeta^\circ = \frac{\bar R_2 \sin(\hat\mu^\circ_2 -2\hat\mu^\circ) }{(1-\bar R)^\frac32}.
\end{equation*}
For symmetric unimodal distributions on a circle, $\hat\zeta^\circ\approx0$. 
\newline
\indent\indent\indent\indent\underline{\bf Circular kurtosis}:
The circular kurtosis is defined~\citep{mardiaandjupp99} as 
\begin{equation*}
\hat\kappa^\circ = \frac{\bar R_2 \cos(\hat\mu^\circ_2 -2\hat\mu^\circ) - \bar R^4 }{(1-\bar R)^2}.
\end{equation*}
For an unimodal angular dataset from the wrapped normal density, $\hat\kappa^\circ\approx0$ and provides an analogue to the usual kurtosis, but for angular observations. 

\begin{figure}[h]
\includegraphics[width=\textwidth]{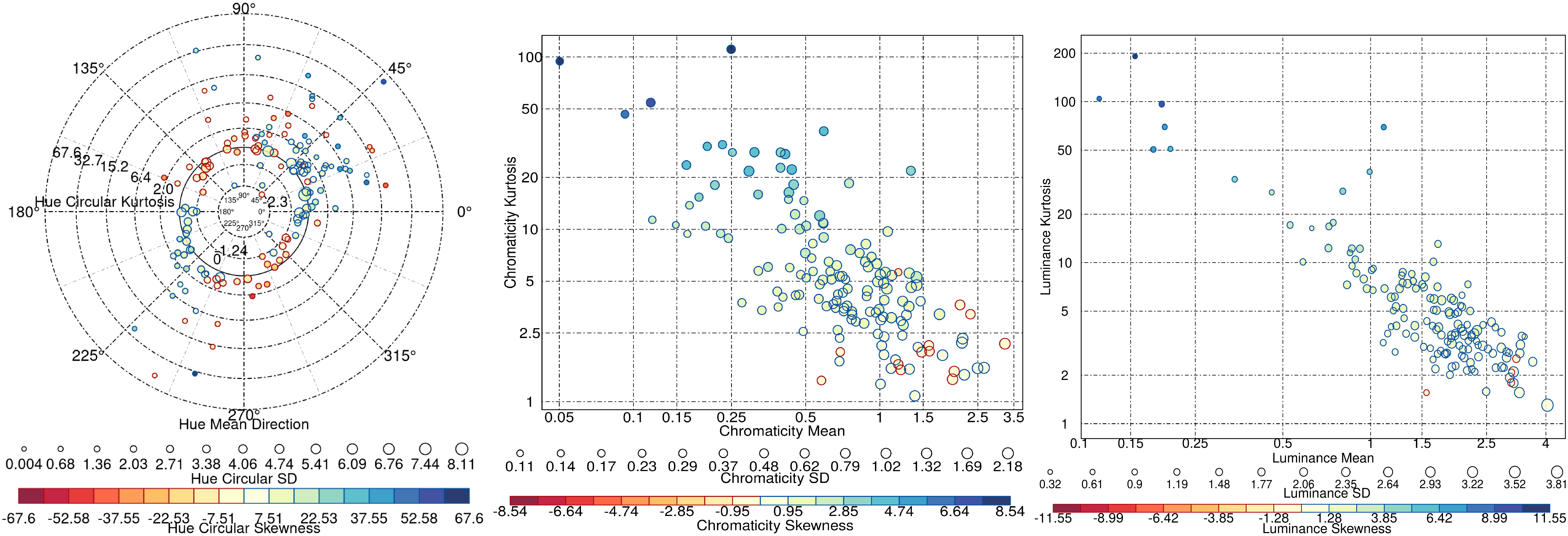}
  \caption{The mean, standard deviation, skewness and kurtosis of the distribution of the hue, chromaticity and luminance in each image. Since hue is an angular measure, its characteristics are in terms of the angle-derived quantities of mean direction $\hat\mu^\circ_H$, circular standard deviation $\hat\sigma^\circ_H$, circular skewness $\hat\zeta^\circ_H$, and circular kurtosis $\hat\kappa^\circ_H$ 
  while chromaticity and luminance are in terms of linear descriptions given by (linear) means $\hat\mu_C$ and $\hat\mu_L$, standard deviations $\hat\sigma_C$ and $\hat\sigma_L$, skewnesses $\hat\zeta_C$ and $\hat\zeta_L$, and kurtosis measures $\hat\kappa_C$ and $\hat\kappa_L$.}
  \label{fig:hclmoments}
\end{figure}
Figure~\ref{fig:hclmoments}
describes the distribution of hue, chromaticity and luminance in each of the 148 images in terms of the circular or linear (as appropriate) mean, SD, skewness and kurtosis. We see that the distributions span a wide gamut in terms of these four important descriptive quantities, and provide reassurance that the images considered in our study span a wide spectrum not just visually and in terms of light conditions or subject matter, but also in terms of the distribution of hue, chromaticity and luminance. We use these measures in the analysis of our results in the next section. 
Further, tagging each image with these distribution characteristics as we have done here also means that we can associate with each desired summary of the hue, chromaticity and luminance distribution a digital image, and our varied set of images provides a ready reference for images with desired distributional characteristics. 
\subsection{Evaluation measure: the visual information fidelity criterion}
The objective of color quantization is to capture all the relevant information while preserving its visual quality, as  far as possible~\citep{yang1998color,huynh2008scope}. The question of answering 
how good one colorspace is at quantizing an image without degrading the image quality has traditionally been settled using the PSNR.
However, the PSNR is directly related to the WCSS, so in our case is not an independent measure. Therefore, we compare the quality of our quantized images with the original in terms of the VIF criterion
\citep{sheikhetal05,sheikhetal06} that  compares the original source image with the quantized image using natural scene
statistics and has been shown to measure the perceptual quality difference between a processed image and its source. 
The VIF measure takes a ratio of the information that can be extracted by the brain from the original and quantized images across multiple subbands of the wavelet decomposition, and takes a value in [0,1], with a VIF value of 1 indicating that
the quantized image is indistinguishable from the original, while a VIF value of 0 means that all of the information in the original image  has been lost in the quantization.

We assessed the quality of the quantized images relative to the original by means of the VIF criterion
\citep{sheikhetal06}. This criterion compares the original source image with the quantized image using natural scene
statistics and is shown to model the perceptual quality difference between a processed image and its source. In particular, the 
VIF is designed to handle images that contain natural scenes, which is the case for many of the images under consideration here. We refer the interested reader to \citet{sheikhetal06} for technical details, which are quite considerable, noting here only that the metric calculates  a ratio of information that could be extracted by the brain from the original and quantized images
across multiple subbands of the wavelet decomposition, and takes a value between 0 and 1. A value of 1 indicates that
the quantized image is indistinguishable from the original, while a value of 0 means that all of the information in the image has been lost during the quantization process.

\section{Results}
\label{sec:results}
This section reports and evaluates the performance of $k$-means color quantization on the RGB, XYZ and LUV  colorspaces. We first report performance on nine representative showcase images of different sizes, and then summarize performance over the other 139 images. Our evaluations are   for images quantized with $k\in \{8,16,32,64\}$ colors, except for the modest-sized {\em statlab} image for which we also use $k\in{128,256}$ in order to get an indication of performance with many colors. 

\subsection{Evaluation on showcase images}
\label{sec:showcase}
We report perfomance using nine showcase images in this section which we briefly describe first. 
\subsubsection{The five moderate to large-sized showcase images}
\begin{figure}[t]
  \centering
  \includegraphics[width=\textwidth]{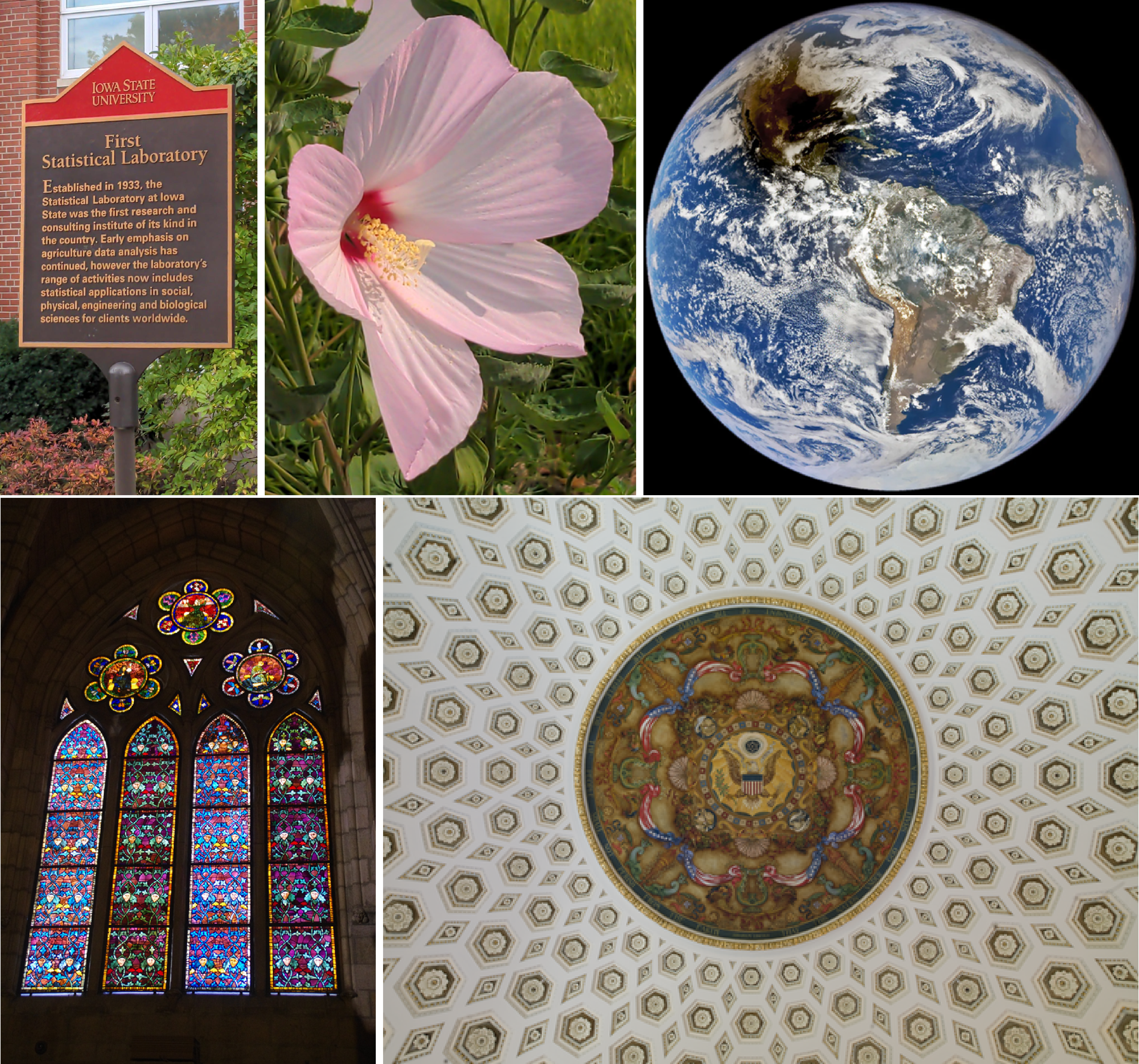}
  \caption{Our showcase images: a modestly-sized image (top left: {\em
      statlab}, of $930{\times}1789$ pixels), two moderate-sized
    images (top center: {\em rosehibiscus}, with $1536{\times}2048$
    pixels, and top right: {\em eclipse}, having $2658{\times}3547$
    pixels) and a fairly large image (bottom: {\em congress} containing
    $5433{\times}7240$ pixels).}
  \label{fig:showcase}
\end{figure}

Figure~\ref{fig:showcase} 
displays the five smaller showcase images used in our investigations. These 
are:
\begin{itemize} 
  \item {\em statlab}: This cropped $486{\times}1155$ digital photograph from 2023 is of  a plaque standing outside the building housing Iowa State    University's Department of Statistics, commemorating the 1933
    establishment there of the first statistical laboratory in the     United States of America. 
  \item {\em rosehibiscus}: This $1536{\times}2048$ digital photograph from July 2023 is of a rose hibiscus in bloom in an Ames, Iowa front yard garden.  
\item {\em eclipse}: This $1976{\times}1647$ digital image, acquired at 16:58 UTC during the October 14, 2023 solar eclipse by NASA’s Earth Polychromatic Imaging Camera imager aboard the Deep Space Climate Observatory satellite, 
shows the umbra from the Moon falling across the southeastern coast of Texas, near Corpus Christi, and is at \url{https://svs.gsfc.nasa.gov/14450}.
\item {\em stainedglass}: This $3264{\times}4928$ digital photograph from August 2019 is of stained glass windows in the nave of the Santa Maria Cathedral of Leon in Spain. 
  \item {\em congress}: This $5433{\times}7240$ image is a view of the
    mural and coffers in the ceiling dome of the Northeast Pavilion in
    the Library of Congress' Thomas Jefferson Building in Washington,
    D.C, and is publicly available from
    \href{http://www.loc.gov/pictures/resource/highsm.11670}{http://www.loc.gov/pictures/resource/highsm.11670}.
  \end{itemize}
\begin{figure}
\centering
  \mbox{\subfloat[The $k$-means color quantized {\em statlab} images in the RGB
  (top row), XYZ (middle row), and LUV (bottom row)
  colorspaces. Color quantization is done with (from left to right) 
  $k\in\{8,16,32,64,128,256\}$.]{\includegraphics[width=\textwidth]{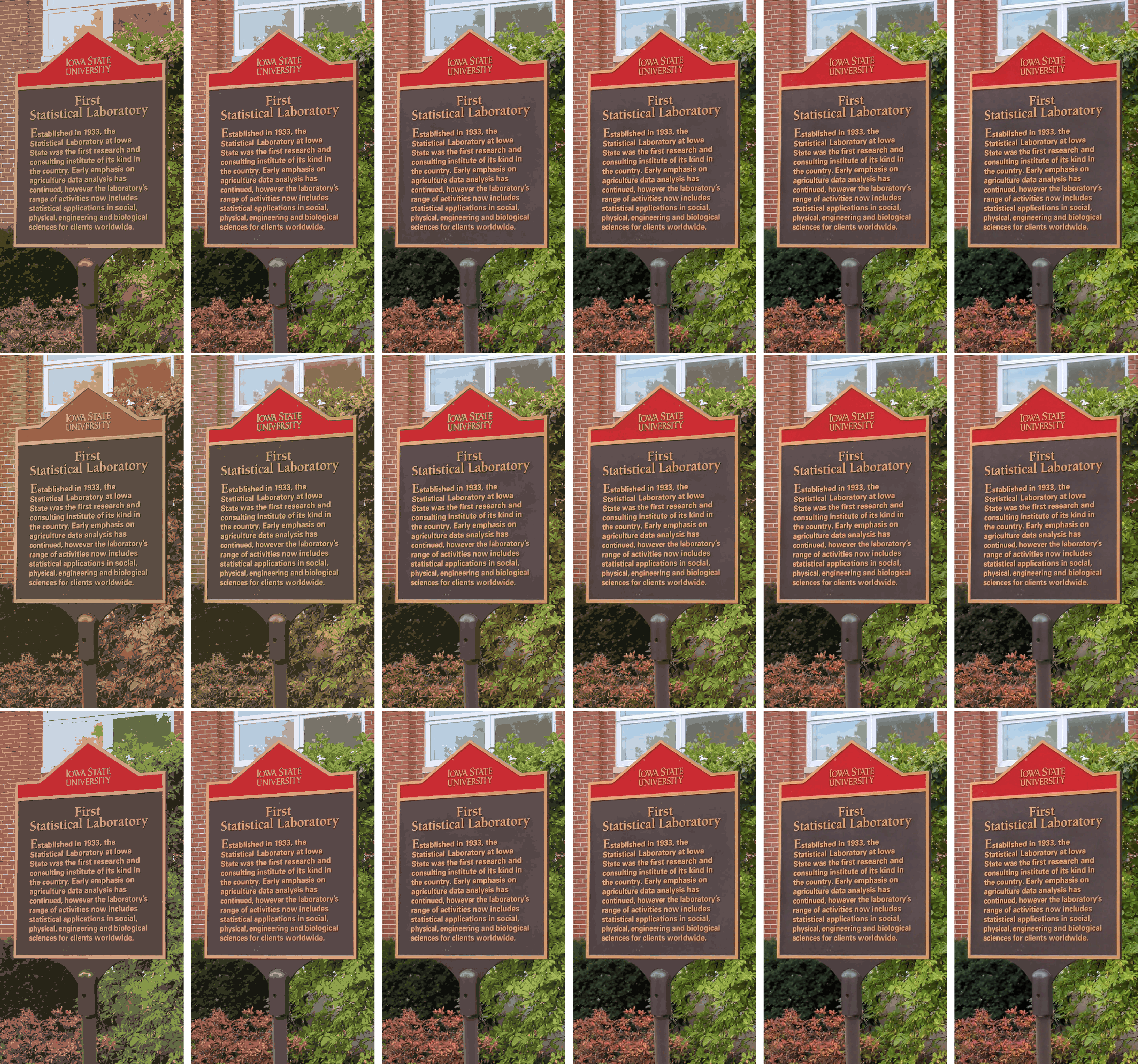}\label{fig:statlab}}}
  \mbox{\subfloat[VIF of the $k$-means color quantized {\em statlab} images to the original.The best performances are highlighted in bold.]{
  \begin{tabular}{|r|cccccc|}\hline
    & \multicolumn{6}{|c|}{$k$} \\ 
Space     & 8 & 16 & 32 & 64 & 128 & 256 \\ \hline
    RGB & 0.500 & 0.607 & 0.713 & 0.795 & 0.858 & 0.900 \\ 
    XYZ  & {\bf 0.589} & {\bf 0.642} & {\bf 0.739} & {\bf 0.831} & {\bf 0.879} & {\bf 0.917}\\ 
    LUV & 0.457 & 0.554 & 0.696 & 0.774 & 0.836 & 0.887\\ \hline
  \end{tabular}
}
  }
\caption{
  Results from $k$-means color quantization of the {\em statlab} image. For each colorspace, the images are   better resolved without increasing $k$. The optimized XYZ colorspace is the best  for all $k$ for this image.}
\label{fig:statlabs}
\end{figure}
Figure~\ref{fig:statlabs}
displays the $k$-means color quantized images for $k\in\{8,16,32,64,128,256\}$, with numerical performance reported in terms of the VIF.  
For $k=8$, the colors  appear to be visually more aligned to the true in the RGB space than in the XYZ space, however, details are better resolved in the XYZ space, and which perhaps is what is driving the superior performance of XYZ in terms of the VIF. In terms of the VIF, $k$-means color quantization in the XYZ space is the best for all $k$. 
This being a smaller image, it is
computationally more practical to investigate performance for high $k$ and
therefore we go up to a larger value for $k$ here than in our other
experimental evaluations. 
We see that improvement in performance is similar with increasing numbers of colors at the higher end, as it is at the moderate end. 
We now discuss performance on the other four images introduced in this section.

\begin{figure}
\includegraphics[width=\textwidth]{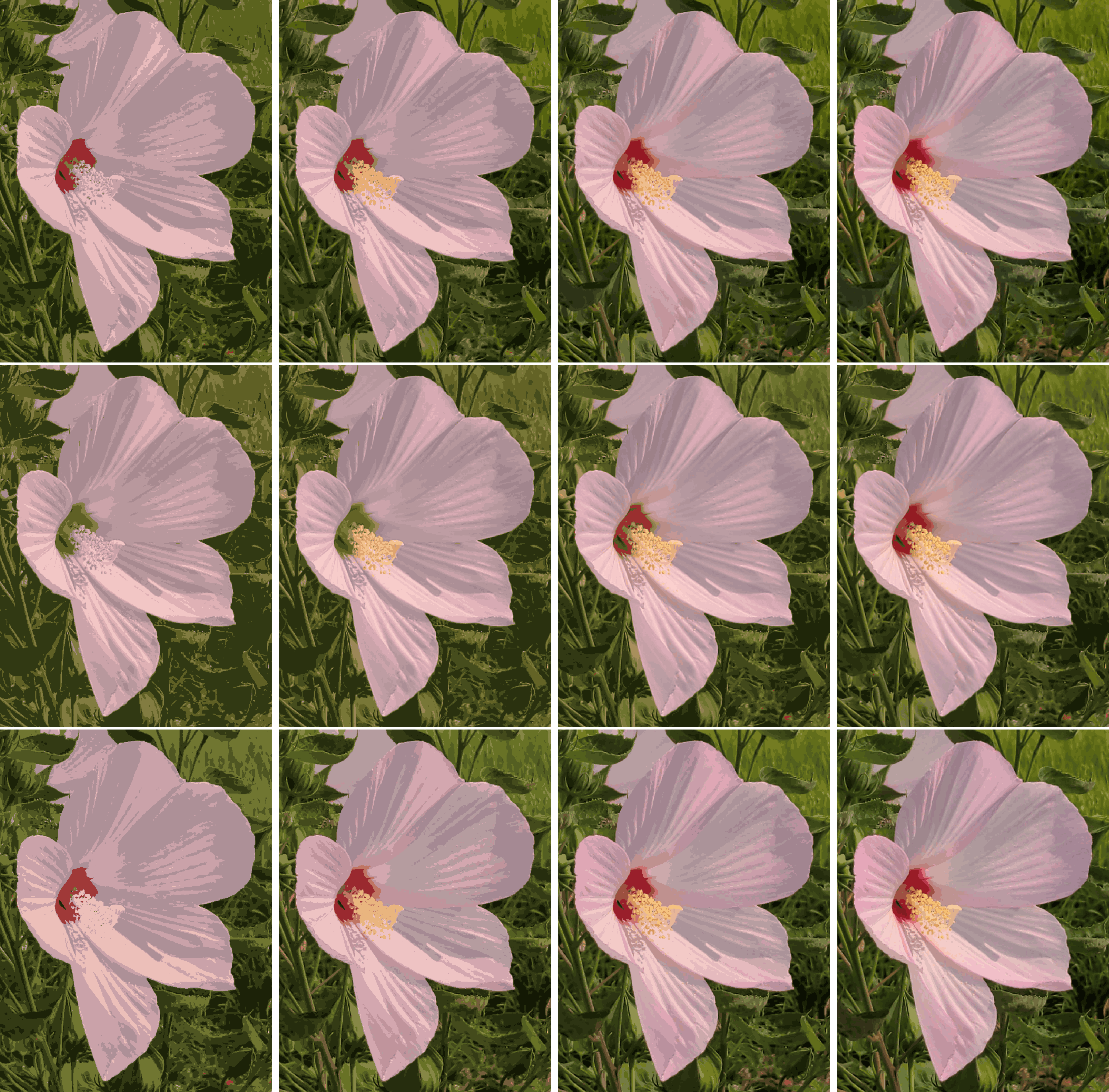}
  \caption{The $k$-means color quantized {\em rosehibiscus} images in the RGB 
  (top row), XYZ (middle row), and LUV (bottom row) spaces, for $k\in \{8,16,32,64\}$ (from left to right).}
\label{fig:rosehibiscus}
\end{figure}

Figure~\ref{fig:rosehibiscus} 
displays performance of $k$-means color quantization of the {\em rosehibiscus} image with different $k$ on the three colorspaces. For $k\in\{8,16\}$, we see that performance is the best on the RGB space, however, it appears that for $k\in\{32,64\}$, performance is best in the XYZ space. This finding is also borne out by the VIF measures in Table~\ref{tab:next4}.
For the {\em eclipse}, {\em stainedglass} and {\em congress} images, the VIFs are a mixed bag, with performance sometimes better in the RGB space and sometimes in the XYZ space.
We now investigate performance on four ultra-large images that we introduce next.

\subsubsection{The four ultra-large images}
\begin{figure}[h]
  \centering
  \includegraphics[width=\textwidth]{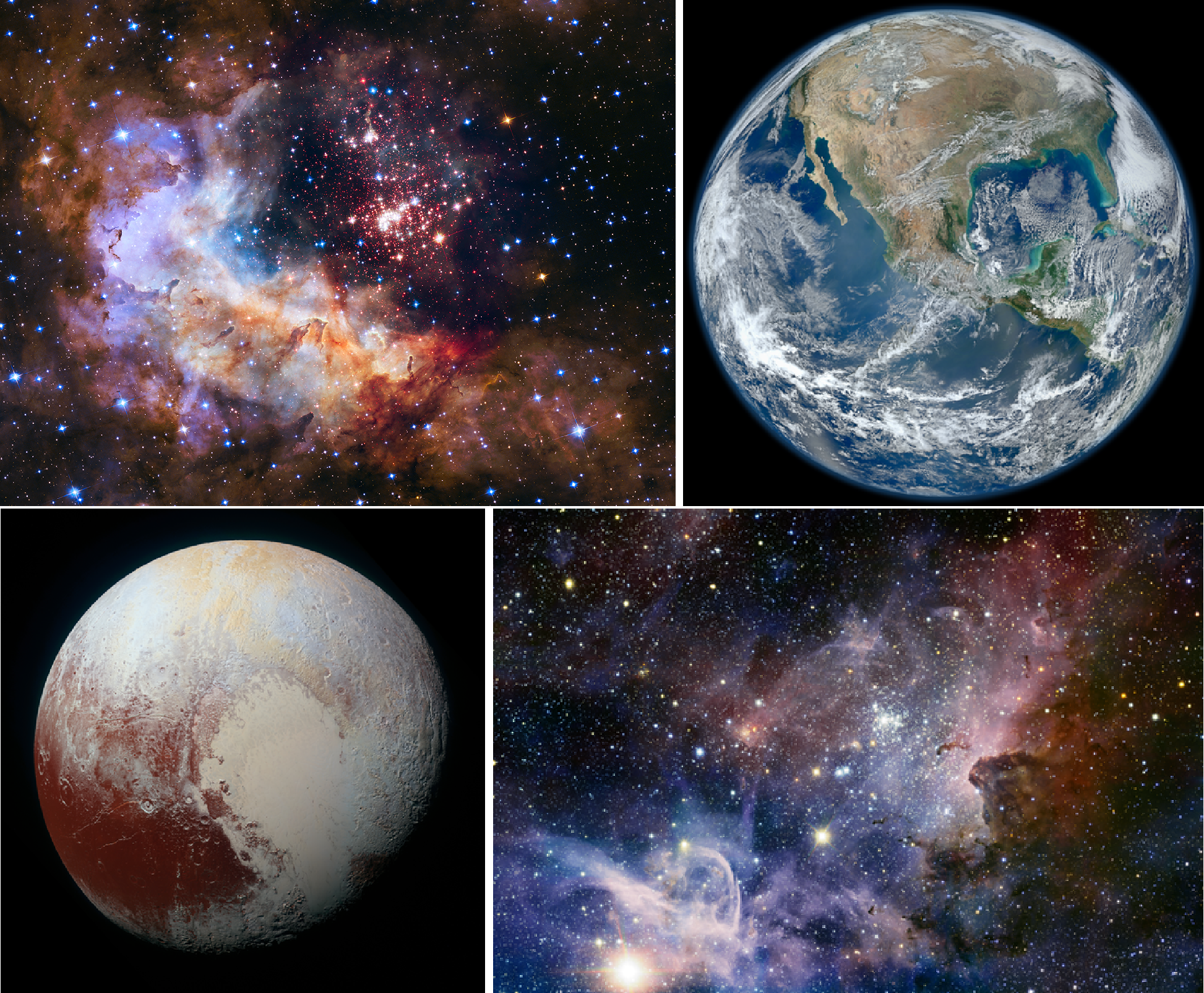}
  \caption{The four ultra-large images: {\em Westerlund2}, {\em earth-HD}, {\em pluto}  and {\em eso1208a}.}
  \label{figsetbig}
\end{figure}

  Figure~\ref{figsetbig}
shows reduced versions of the four largest images that we considered and also used in our showcase evaluations. They are:
  \begin{itemize}
	  \item {\em Westerlund2}: This $8919{\times}6683$ image, taken by NASA's Hubble telescope in its 25th year in 2015, is of a giant cluster of about 3,000 stars called Westerlund 2, named for the Swedish astronomer Bengt Westerlund, who discovered the grouping in the 1960s. The cluster resides in a raucous stellar breeding ground known as Gum 29, located in the Carina constellation 20,000 light-years away from Earth. 
	  \item {\em earth-HD}: This $8000{\times}8000$ image of our Blue Planet was released by NASA on 24th January 2012 and is created from photographs taken by the Visible/Infrared Imager Radiometer Suite (VIIRS) instrument on board the Suomi NPP satellite. It provides one of the most detailed images of the earth that has ever been created.
	  \item {\em pluto}: This $8000{\times}8000$ high-resolution color view of Pluto was captured by NASA’s New Horizons spacecraft on July 14, 2015 and combines blue, red and infrared images taken by the Ralph/Multispectral Visual Imaging Camera. 

	  \item {\em eso1208a}: This broad $8926{\times}13092$ image of the Carina Nebula, a region of massive star formation in the southern skies, was taken in infrared light using the HAWK-I camera on the European Soutern Observatory’s Very Large Telescope, and shows many previously hidden features scattered across a spectacular celestial landscape of gas, dust and young stars.
\end{itemize}
\begin{table}[h]
  \caption{VIF of  $k$-means color quantized images with the true {\em rosehibiscus}, {\em eclipse}, {\em stainedglass} and {\em congress} images. The best performance for each setting is highlighted in bold.
  \label{tab:next4}}
  \centering
  {\small 
  \begin{tabular}{|r|ccc|ccc|ccc|ccc|}\hline
    & \multicolumn{3}{|c|}{\em rosehibiscus} & \multicolumn{3}{|c|}{\em eclipse} & \multicolumn{3}{|c|}{\em stainedglass} & \multicolumn{3}{|c|}{\em congress} \\ \hline
    $k$ & RGB & XYZ & LUV & RGB & XYZ & LUV & RGB & XYZ & LUV & RGB & XYZ & LUV \\ \hline 
    8& {\bf 0.462} & {\bf 0.462} & 0.432 & 0.552 & {\bf 0.601} & 0.524 & 0.290 & {\bf 0.294 } & 0.286 & {\bf 0.632} & 0.618 &  0.612 \\
    16 &  {\bf 0.597} & 0.596 & 0.550 & {\bf 0.697} & 0.674 & 0.647 & {\bf 0.511} & 0.370 & 0.478 &  0.726 & {\bf  0.788} & 0.675\\
    32 & 0.695 & {\bf 0.738} & 0.662 & {\bf 0.799} & 0.767 & 0.765 & {\bf 0.624} & 0.436 & 0.596&0.818 & {\bf 0.840} & 0.780\\
    64 & 0.768 & {\bf 0.830} &  0.746 & {\bf 0.866} & 0.843 & 0.846 & {\bf 0.715} & 0.609 & 0.696 &  0.881 & {\bf 0.897} & 0.850\\ \hline
  \end{tabular}}
\end{table}
\begin{table}[h]
  \caption{VIF of  $k$-means color quantized images with the true {\em Westerlund2}, {\em earth-HD}, {\em pluto} and {\em eso1208a} images. The best performance for each setting is highlighted in bold.
  \label{tab:ultra}}
  \centering
  {\small
  \begin{tabular}{|r|ccc|ccc|ccc|ccc|}\hline
     & \multicolumn{3}{|c|}{\em Westerlund2} & \multicolumn{3}{|c|}{\em earth-HD} & \multicolumn{3}{|c|}{\em pluto} & \multicolumn{3}{|c|}{\em eso1208a} \\ \hline
    $k$ & RGB & XYZ & LUV  & RGB & XYZ & LUV  & RGB & XYZ & LUV  & RGB & XYZ & LUV \\ \hline
    8 & {\bf 0.412} & 0.308 & 0.399 & {\bf 0.628} & 0.606 & 0.557 & {\bf 0.468} & 0.447 & 0.443 & 0.504 & 0.426 & {\bf 0.520}\\
    16 & {\bf 0.536} &  0.464 & 0.512 & 0.723 & {\bf 0.756} & 0.710 & {\bf 0.640} & 0.637 & 0.577&  0.616 & 0.523 & {\bf 0.646}\\
    32 & {\bf 0.660} & 0.567 & 0.650 & 0.853 & {\bf 0.862} & 0.831 & {\bf 0.780} & 0.712 & 0.749&0.724& 0.652 & {\bf 0.736}\\
    64 & 0.730 & 0.674 & {\bf 0.752} & 0.906 & {\bf 0.907} & 0.886 & {\bf 0.873} & 0.765 & 0.861&
    0.799 & 0.736 & {\bf 0.813}\\ \hline
  \end{tabular}
  }
\end{table}
Table~\ref{tab:ultra}
shows the performance of $k$-means color quantization for these images in the three colorspaces. The best performance is in different colorspaces according to both $k$ and the image, for we see that in some cases, the RGB space is where $k$-means color quantization works best for certain $k$, while for others, the XYZ space works best at some other $k$, and indeed, for some other $k$ and images, the LUV space is the winner.
\subsection{Evaluation in a large-scale study}
  \label{sec:large-scale}
  The results of our investigations in Section~\ref{sec:showcase} do not provide a particularly clear picture on the colorspace where $k$-means color quantization does best. Therefore, we investigate performance of $k$-means color quantization by also including the 139 additional images that are briefly  described in Appendix~\ref{app:addlimages}.
\begin{figure}[h]
  \includegraphics[width=\textwidth]{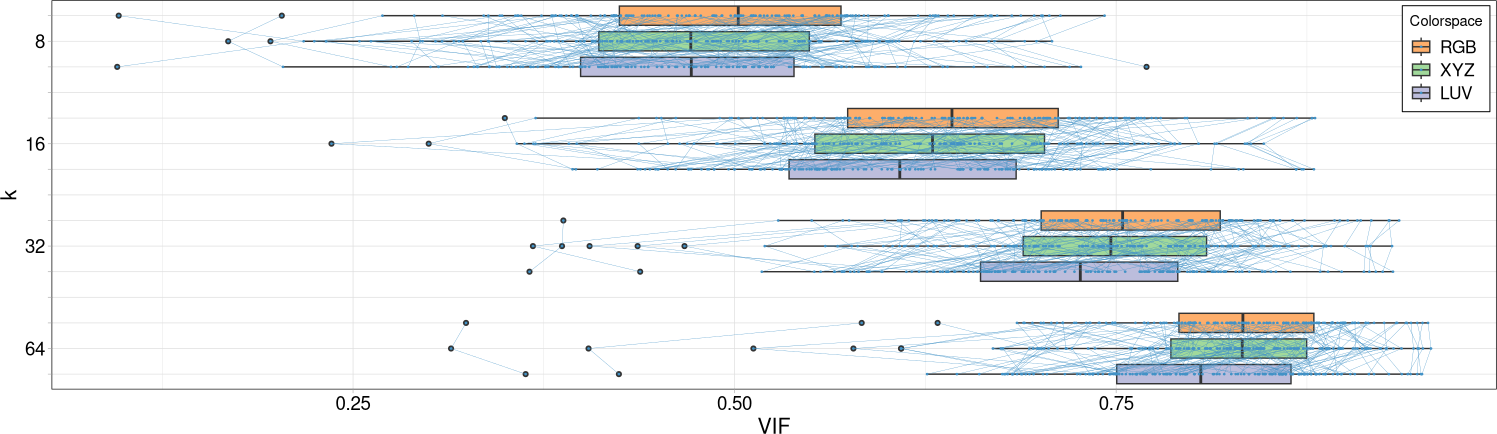}
  \caption{The distribution of the VIFs of the $k$-means color quantized images done on the three colorspaces for $k=8,16,32,64$. The lines in the linked boxplots indicate the individual VIFs and allow us to track the performance of the same image upon $k$-means color quantization across the different colorspaces and for a specified $k$.}
  \label{fig:vifs}
  \end{figure}
  Figure~\ref{fig:vifs}
displays the  VIFs obtained for the 148 images after using $k$-means color quantization in the three colorspaces. Expectedly, the VIFs increase with increasing $k$ in all the three colorspaces. However, which colorspace is better for $k$-means color quantization varies from image to image. In general, it appears that the distribution of the VIFs is higher for the RGB-space quantized images than for the images quantized in the XYZ and the HCL colorspaces, and in that order. 
\begin{table}
  \caption{The number of cases for which $k$-means color quantization is best in each colorspace for $k=8,16,32,64$.}
   \label{tab:freq.performance}
  \centering
  \begin{tabular}{|c|cccc|}\hline 
    & \multicolumn{4}{|c|}{$k$} \\ 
    Colorspace & 8 & 16 & 32 & 64\\ \hline
    RGB &  78 &   73 &  74 &   73\\
XYZ  &   49 &  58 &  66 &  67\\
LUV &   21 &  17 &   8 &   8 \\ \hline
  \end{tabular}
\end{table}
Table~\ref{tab:freq.performance} displays the number of cases 
where $k$-means color quantization does best in each of the colorspaces, for $k=8,16,32,64$. In general, it appears that RGB is the best in about half the cases for all $k$. 
For the other half, it appears that XYZ is better than LUV in more cases, with the gap increasing substantially with larger $k$. 

We refer back to Figure~\ref{fig:hcldist}
that uses palette color to characterize the best performing colorspace for $k=64$   in terms of the distributions of hue, chromaticity and luminance. It appears that the hue is more spread out for larger images when the RGB space hosts the better $k$-means color quantization performer over the XYZ space. Further, for small images, it appears that the hues are far more concentrated than for larger images, but the mean directions of these distributions are generally in different directions for similar-sized images where the different colorspaces do better. Similar statements can also be made on the distribution of chromaticity and luminance. However, these observations are only qualitative so we now attempt to provide greater understanding of the images when each method performs better by a more thorough quantitative analysis.

\subsubsection{Statistical analysis of our results} 
The goal of our statistical analysis is to understand the performance of $k$-means color quantization in the different colorspaces. Our response variable is obtained from the VIF measure of each $k$-means color quantized image relative to the true, for $k=8,16,32,64$. The VIF is a measure between 0 and 1, so we use a logit transformation 
\begin{equation}
  \lambda_{\mbox{VIF}_{s,k}} = \log\left(\frac{\mbox{VIF}_{s,k}}{1-\mbox{VIF}_{s,k}}\right)
  \label{eq:logit}
\end{equation}
that converts the measure to be on the real line $\R$. Here $\mbox{VIF}_{s,k}$ represents the VIF for the $k$-means color quantized image obtained in the $s$th colorspace, for $k\in \{8,16,32,64\}$ and $s\in\{\mbox{RGB, XYZ, LUV}\}$. The logit transformation is a bijective monotone transformation, so it does not change the direction of performance of VIF in any of the colorspaces.  Further, we set the baseline for each $k$ and each image to be performance of $k$-means color quantization in the RGB space. As a result, we explain performance in terms of the difference in logit VIFs between $k$-means color quantization in the XYZ and the RGB spaces and the difference in logit VIFs between $k$-means color quantization in the LUV and RGB spaces. Formally, our modified response variable for the $ith$ image is a $2{\times}4$ matrix $\by_i = ((y_{i,\ell,k}))$ where 
\begin{equation}
  y_{i,\ell,k} = \lambda_{\mbox{VIF}_{\ell,k}} - \lambda_{\mbox{VIF}_{\mbox{RGB},k}},\quad  \mbox{for $k\in \{8,16,32,64\}$ and $\ell\in\{\mbox{XYZ, LUV}\}$.}
  \label{eq:response}
\end{equation}
We model this response in terms of the size of the image as well as the image-descriptive variables that characterize the distribution of hue, chromaticity and luminance. The image size is provided in terms of the number of pixels provided as a pair $(I,J)$ where $I$ and $J$ denote the number of pixels in the shorter and longer edges of the image.
The image-descriptive variables  are the four measures that we have chosen to characterize each of the hue, chromaticity and luminance distributions. All told, we consider for inclusion $I$, $J$,
$\hat\sigma^\circ_H$, $\hat\zeta^\circ_H$, $\hat\kappa^\circ_H$, 
$\hat\mu_C$, $\hat\sigma_C$, $\hat\zeta_C$, $\hat\kappa_C$, $\hat\mu_L$, $\hat\sigma_L$, $\hat\zeta_L$ and $\hat\kappa_L$ as covariates in our analysis. Therefore, we have 14 potential features that can explain differences in the VIFs (after logit transformation) in the $k$-means color quantized images underlying the XYZ- and LUV-spaces, vis-a-vis that obtained in the RGB-space. 

We reiterate that the goal of our analysis is to understand the conditions for the performance of $k$-means rather than prediction. This is the reason that we eschew deep learning algorithms that excel at prediction but provide models that are not easily interpreted. At the same time, our model needs to be flexible in order to account for the complexity in the predictors and the responses. These complexities include the facts that the response is multivariate (indeed matrix-variate), that $\mu^\circ_H$ is an angular quantity, or that there is not necessarily a linear, or for that matter, a specific prescribed functional relationship between the response matrix and the predictors. Additionally, the error in the relationship may not be from a specific class of distributions. It is for this reason that we employ the nonparametric technique of multivariate regression trees (MVRTs)~\citep{segal92,death02} that extends to the multivariate response setting, tree-structured (or recursive partitioning) regression. A 
major  benefit of tree-structured methods is its ability to provide interpretable prediction rules 
through a natural, recursive approach by subdividing data into homogeneous subgroups. 
For continuous responses, simple (terminal) subgroup summaries (typically means) are the predictions. Similar to the case with univariate responses, tree-based ensemble methods called multivariate random forests (MVRFs) can be used to borrow strength from correlated features in sparse settings~\citep{segalandxiao11}. Features in a tree are selected on the basis of its importance through variable importance measures (VIMs) that score each feature according to the strength of the model's dependence on that variable. Recent work~\citep{sikdaretal25} provides VIMs for MVRFs based on its ability to achieve split improvement defined as the difference in the multivariate responses between the two children at each parent node. A benefit of these VIMs is that they allow us to study the strength of dependence both globally and on a per-response basis~\citep{sikdaretal25}. We use these VIMs within MVRFs to study the importance of each of our 14 features in explaining $(y_{\cdot,\mbox{XYZ},k}, y_{\cdot,\mbox{LUV},k})$ for $k=8,16,32,64$.

\begin{figure}
  \includegraphics[width=0.99\textwidth]{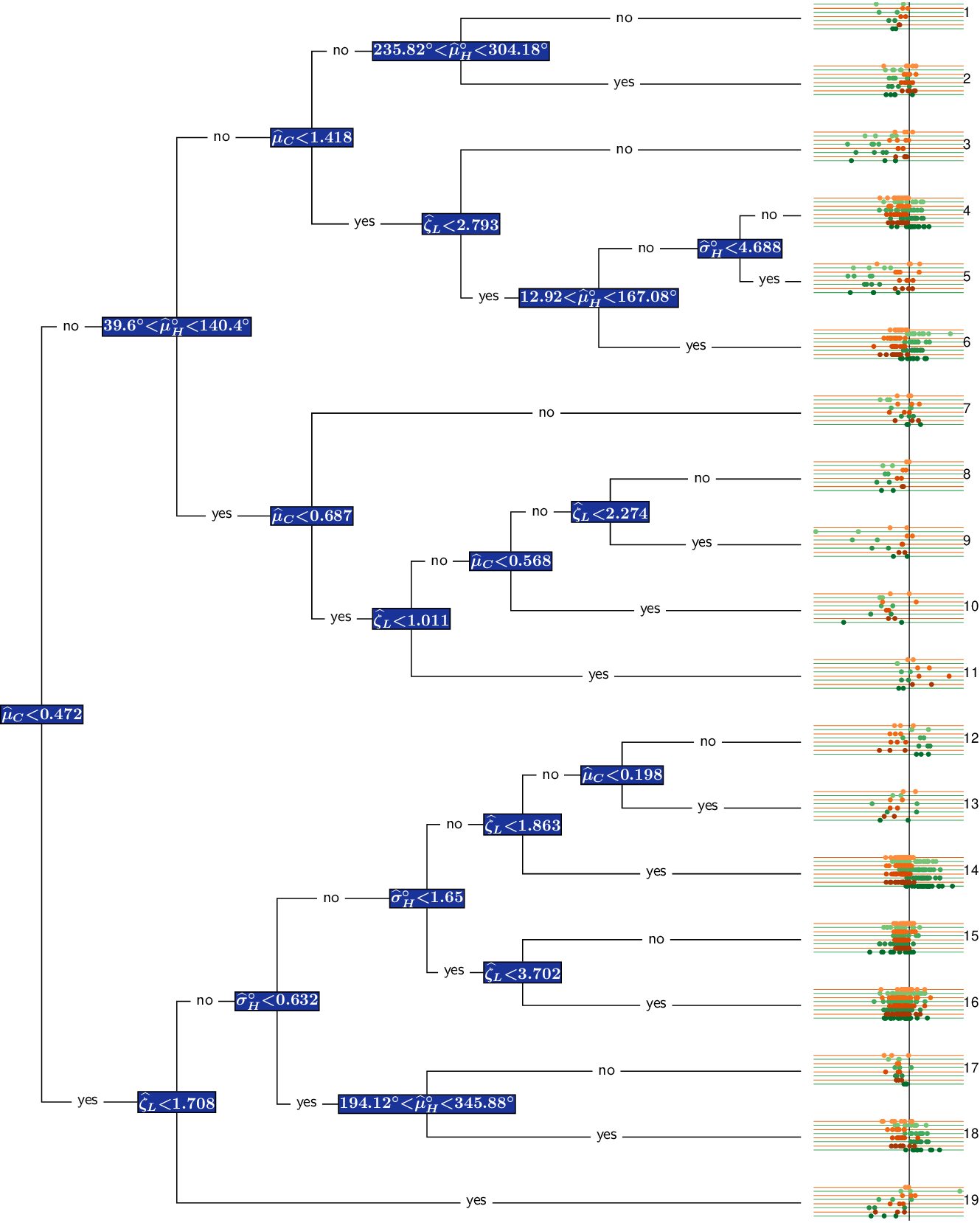}
  \caption{The MVRT upon fitting  $\hat\mu_C$, $\hat\mu^\circ_H$, $\hat\zeta_L,\hat\sigma^\circ_H$ to the matrix-valued $((y_{i,\ell,k}))$s as defined in \eqref{eq:response}. The leaves at each terminal node (labled 1--19 for ready reference) display the matrix-valued responses assigned to that node, with greens and oranges representing the   $y_{i,\mbox{XYZ},k}$ and $y_{i,\mbox{LUV},k}$ values for the $ith$ digital image, and darker shades indicate higher $k\in \{8,16,32,64\}$. The vertical line through all the data displays is at zero: green/orange  observations to the right/left of this line are where $k$-means color quantization does better/worse in XYZ/LUV space than in RGB space.} 
  \label{fig:mvpart-tree}
\end{figure}

Table~\ref{tab:vims}
provides the features corresponding to the top ten VIMs. From the table, it is clear that $\hat\mu_C$ is the leading variable
\begin{table}[h]
  \caption{The features with the top ten VIMs in decreasing order.}
  \label{tab:vims}
  \begin{tabular}{c|ccccccccccc}\hline
    { Feature} & $\hat\mu_C$ & $\hat\mu^\circ_H$ & $\hat\zeta_L$ & $\hat\sigma^\circ_H$ & $\hat\mu_L$ & $\hat\zeta_C$ & $\hat\zeta^\circ_H$ & $\hat\sigma_C$ & $\hat\sigma_L$ &$\hat\kappa_L$\\ \hline
    { VIM} & 29.018 & 17.464 & 14.538 & 12.461 & 11.226 & 9.536 & 9.374 & 9.364 & 8.311 & 7.233\\ \hline 
  \end{tabular}
\end{table} 
\begin{table}[h]
  \caption{Cross-validated error of the fitted MVRT as each feature with decreasing VIM is included in  the model.}
  \label{tab:models}
  \begin{center}
  \begin{tabular}{c|ccccccc}\hline
    {\bf Added feature} & $+\hat\mu_C$ & $+\hat\mu^\circ_H$ & $+\hat\zeta_L$ & $+\hat\sigma^\circ_H$ & $+\hat\mu_L$ & $+\hat\zeta_C$ & $+\hat\zeta^\circ_H$ \\ \hline
    {\bf Model CV error } & 1.0775 & 1.0157 & 1.0138 & 0.9409 & 0.9515 & 0.9997 & 1.0217 \\ \hline 
  \end{tabular}
  \end{center}
  \end{table} 
in the construction of a MVRT ensemble, followed at a distance by $\hat\mu^\circ_H$, which is turn is followed at a modest distance by $\hat\zeta_L$, and $\hat\zeta_L$,  $\hat\sigma^\circ_H$, $\hat\mu_L$  and so on, in that order. We note that the VIM values have some uncertainty in them, and while the exact values change from the application of one random forest to another, the ordering of the features in terms of their VIMs is remarkably stable and did not pretty much change in any of  100 runs.
The VIMs  provide us with a measure of the contribution of each feature in the regression tree, but not the actual tree itself, so we build 
increasingly richer trees by progressively including covariates in the order of their decreasing VIMs.
Table~\ref{tab:models} provides the minimum cross-validated (CV) error over 100 runs for each of the MVRTs obtained by progressively including variables in the tree in the decreasing order of their VIMs. We see that the lowest CV error rate is obtained when the tree has the four explanatory features $\hat\mu_C$, $\hat\mu^\circ_H$, $\hat\zeta_L$ and $\hat\sigma^\circ_H$. There are nineteen terminal nodes on this tree, and 18 splits, as seen in Figure~\ref{fig:mvpart-tree} which displays the tree obtained from the above fit. 
The leaves at each terminal node represent the data at each node. Since each response is 
matrix-valued, Figure~\ref{fig:mvpart-tree}
uses color and shading to display the colorspace and $k$ in the $y_{i,\ell,k}$ of \eqref{eq:response} to represent the responses assigned to each node. We now discuss these results to better inform us as to when $k$-means color quantization works better or worse in a particular colorspace than in the others.

The first (as per the labels of Figure~\ref{fig:mvpart-tree}) terminal node is for images with $\hat\mu_C{>}1.418$ and $\hat\mu^\circ_H {\in} [0^\circ,39.6^\circ]\cup[140.4^\circ,235.82^\circ]\cup[304.18^\circ,360^\circ],$ and where we see $k$-means color quantization produce better results in the RGB space than in the LUV and XYZ spaces, in that order. Here, $k$ does not matter that much with regard to RGB vis-a-vis LUV, but even then  quantized images in the XYZ space are sometimes relatively worse for smaller $k$ than larger $k$. A somewhat similar ordering in performance, but with relatively different  values, is reported in the third, eighth, ninth, and tenth terminal nodes. The third  node has images with $\hat\zeta_L {\geq}2.793$, $\hat\mu_C{\in}[0.472,1.418)$ and $\hat\mu^\circ_H{\not\in}(39.6^\circ, 140.4^\circ)$, and substantial positively-skewed luminance, having $\hat\zeta_L \geq2.793$), while the eighth, ninth and tenth terminal nodes contain images that are characterized with high ($\hat\zeta_L {\geq}2.274$, in the eighth node) to moderate  ($\hat\zeta_L{\in} [1.011,2.274)$, in the ninth or tenth nodes) skew in luminance, $\hat\mu^\circ_H{\in}(39.6^\circ,140.4^\circ)$, and either $\hat\mu_C{\in} [0.568, 0.687)$ (for images in the eighth and ninth nodes) or $\hat\mu_C{\in} [0.472, 0.568)$ (for images belonging to the tenth terminal node). 
While  the seventeenth terminal node shows similar performance, performance in the XYZ space is sometimes better than in the LUV space (but bested in all but one case by performance in the RGB space). The images in this terminal node have at least moderately high skew in  luminance ($\hat\zeta_L{\geq}1.708$),  small $\hat\mu_C$ of below 0.472 and hue circular SD $\hat\sigma^\circ_H{<}0.632$, and $\hat\mu^\circ_H{\in}(194.12^\circ,345.88^\circ)$.

The second terminal node has images with a similar set of $\hat\mu_C$ as those assigned to the first node, but with $\hat\mu^\circ_H{\in}(235.82^\circ, \allowbreak 304.18^\circ)$. For these images, LUV-space is sometimes where $k$-means color quantization performs the best, while the XYZ space is (with the exception of one $k{=}64$ case) worse for $k$-means color quantization than the RGB space. Somewhat similar ordering but different relative performance,  is also reported in the images in the  fifth terminal node that  contains images whose hues have modest circular SD ($\hat\sigma^\circ_H {<} 4.688$), $\hat\mu^\circ_H{\not\in} (12.92^\circ,167.08^\circ)$, low- to moderate-skewed luminance with $\hat\zeta_L{<}2.793$, and moderate mean chromaticity of $\hat\mu_C{\in}[0.472,1.418)$. 
The nineteenth node, with images that have lower mean chromaticity ($\hat\mu_C{<}0.472$) and luminance skewness ($\hat\zeta_L{<}1.708$) also has similar results, but here the performance in XYZ space is relatively much worse than in LUV space. 
The seventh node, comprising images with  mean hue $\hat\mu_H^\circ$ between $39.6^\circ$ and $140.4^\circ$ and higher average chromaticity ($\hat\mu_C{\geq}0.687$) is also somewhat similar, but here, performance in XYZ colorspace is several times the best for higher $k$. 

The fourth, sixth, twelfth, thirteenth, fourteenth, and perhaps even the eighteenth terminal nodes are for images where $k$-means color quantization in the XYZ space is, in many cases, the clear winner (and indeed in a few cases in the twelfth and fourteenth nodes, performance in even the LUV space is better than in the RGB space). Images assigned to the fourth and sixth terminal nodes have  moderate $\hat\mu_C{\in}[0.472,1.418)$, lower-skewed luminance ($\hat\zeta_L {<} 2.793$), and $\hat\mu^\circ_H{\not\in}(12.92^\circ,167.08^\circ)$, but for the fact that images in the fourth node also have higher hue circular SD ($\hat\sigma^\circ_H {\geq} 4.688$). Images in the twelfth, thirteenth and fourteenth terminal nodes are characterized by moderate to low $\hat\mu_C$ of less than $0.472$, moderate to higher skew in luminance $(\hat\zeta_L{\geq}1.708)$ and high hue circular SD $(\hat\sigma^\circ_H{>}1.65)$. There is greater uncertainty, for lower $k$, in whether $k$-means color quantization does better in the XYZ than RGB space, in the fourteenth node (where, additionally, the skew in luminance is moderate at $\hat\sigma_L{<}1.863$), but where performance in LUV space is at best marginally better than in RGB space, but far more often marginally to substantially worse.  
On the contrary, for images with higher skew in luminance ($\sigma_L{\geq}1.863$), the distinctions in performance are emphatic (with performance in XYZ space much better than in RGB, and performance in the RGB space  almost always besting that in  LUV space) in the twelfth node with higher $\hat\mu_C$ of at least 0.198. But when $\hat\mu_C{<}0.198$, as in the thirteenth terminal node images, then we see that performance in XYZ space relative to that  in RGB space is more erratic, but here also the differences in performance are emphatic (either way), and only for one $k{=}8$ case does LUV come out on top.

The eleventh terminal node images, having moderate  average chromaticity ($0.472{\leq}\hat\mu_C{<}0.687$), lower luminance skewness ($\hat\zeta_L{<}1.011$) and hue mean direction of between $39.6^\circ$ and $140.4^\circ$ is where LUV-space performance is decidedly the best, or at least moderately so, and where performance in the XYZ space is at least moderately worse than in the RGB space. 

The remaining two terminal nodes are for images with lower average chromaticity ($\hat\mu_C{<}0.472$), moderate hue circular SD ($0.632{\leq}\sigma^\circ_H{<}1.65$) and for moderate and large luminance skewnesses. For those images that additionally have $\hat\zeta_L{\geq}3.702$ (fifteenth terminal node), performance of $k$-means color quantization in the RGB space is in many  cases the best, while that in the  XYZ space is generally the worst while images in the sixteenth terminal node, with moderate luminance skewnesses ($1.708{\leq}\hat\zeta_L{<}3.702$), are less clear in terms of performance for smaller $k$, but RGB-space is increasingly the clear winner as $k$ increases. For these sixteenth node images, honors are even between the XYZ and LUV spaces, with not much distinction even for different $k$.

The results of our analysis provide a nuanced and deeper understanding of the characteristics of images  for  which $k$-means color quantization performs well in the machine-based RGB colorspaces or in the  more human perception-based XYZ and LUV colorspaces. Our assessment has been in terms of the distributional characteristics of hue, chromaticity and luminance, an approach that is novel but can easily be extended to understand performance of image processing algorithms in other settings. 

\section{Discussion}
This paper analyzes, in a large-scale study, the use of machine-based RGB and human perception-based XYZ and LUV (equivalently HCL) colorspaces for the performance of $k$-means color quantization of digital images. 
A few authors~\citep{yoonandkweon04,berlinskiandmaitra25} have shown that performance of  color quantization may be improved by considering human perception-based colorspaces, but their investigations have been on smaller studies of only a few images. The study here is on a large number (148) of images that is not only a larger testbed of images than what has  typically been investigated in the past, but is also very varied and expansive in terms of image sizes, scenes, subjects, lighting conditions and settings, with care also taken to characterize and understand the differences in these settings.  The VIF measure~\citep{sheikhetal05,sheikhetal06} was used to evaluate the quality of the quantized image relative to the original. Our investigations here show that the machine-based RGB colorspace can hold its  own with regard to $k$-means color quantization, and indeed is the better performer in about  half the cases. For the other cases, we see that at lower quantization levels ($k$), $k$-means color quantization in  LUV space is often the better  performer, but certainly not at the same level as that of $k$-means  color quantized images in XYZ space: indeed the difference only increases with  $k$. Our investigations here also provide understanding on when each colorspace is the better performer. In order to do so, our analysis specifically characterizes each image in terms of the distribution of the hue, chromaticity and luminance at its pixels, and then summarizes these distributions in terms of circular (for hue) and linear (for chromaticity and luminance) statistics, in particular, in terms of their (circular, for hue) means, SDs, skewness and kurtosis. These variables, along with the image sizes, are included in a multivariate random forest~\citep{segalandxiao11} framework to assess, through variable importance measures~\citep{sikdaretal25}, the distributional characteristics that are then included in a multivariate regression tree~\citep{segal92,death02} to provide understanding of when color quantization does well in which colorspace. (Despite being included, image sizes are not picked up as important variables to use in our multivariate regression tree fits.) Figure~\ref{fig:mvpart-tree}
provides us with a detailed perspective on the characteristics of images that informs us whether for any given image, we should perform $k$-means color quantization in the machine-based RGB or in one of the human perception-based XYZ or LUV (indeed HCL) colorspaces. 

Our investigations in this paper have been limited to $k$-means color quantization of digital images. However, as mentioned, for instance, in a recent review~\citep{celebi23}, there exist other color quantization methods that can also be evaluated for performance in other colorspaces. For instance, one may consider the use of model-based methods for color quantization that borrow ideas from model-based clustering and are a more principled alternative to methods such as $k$-means. Also, beyond the specific question of which colorspace should be used for $k$-means color quantization and when, this paper also makes several important and innovative contributions. Our 148 publicly available images are characterized in terms of the distributions of hue, chromaticity and luminance at its pixels, and these distributions are themselves summarized in terms of circular and linear statistics. They provide a quantified image dataset to the scientific community for evaluation and testing. In that sense, it adds for the scientific community  another dataset that has lately been made available for this purpose~\citep{celebietal23}. Note however, the novel quantification of the characteristics of the images in our dataset, that can also be used to numerically describe other images available in the literature.  Our characterization of  images  means that we can pull up images with desired distributional characteristics on the hue, chromaticity and luminance, and evaluate and understand performance in other contexts beyond $k$-means or other color quantization methods. 
Thus, we see that there are a number of avenues that get expanded with the treatment and perspective provided in this paper, highlighting the significance of our approach far beyond the specific context of this paper,  and that may be worthy of attention.





\appendices    

\section{The  images used in our  large-scale study}
\label{app:addlimages}
There were 148 images used in our large-scale study, of which nine have been described in Section~\ref{sec:showcase}. 
We now describe the remaining 139 digital images. These 139 images are arranged in terms of increasing size in terms of total number of pixels and, for convenience, grouped into three sets. Therefore the first set displayed in Fig~\ref{fig:small} contains the first 50 images, the second set in Fig~\ref{fig:medium} contains the next 44 images, and the third set in Fig~\ref{fig:large} contains the remaining 41 images. (Note that these sets are only nominally specified for our convenience in describing and displaying their thumbnails, and that all images are analyzed together.
Further, and only to permit their reproduction here, all (true)
images displayed in this paper are of lower resolution than the
originals. We emphasize that all reported evaluations were done on the original images
at native resolution.)
We now individually describe the images that form these three sets in increasing order.
\subsection{The first set of images}
\begin{figure}                                                  
\centering                                                   
\includegraphics[width=\textwidth]{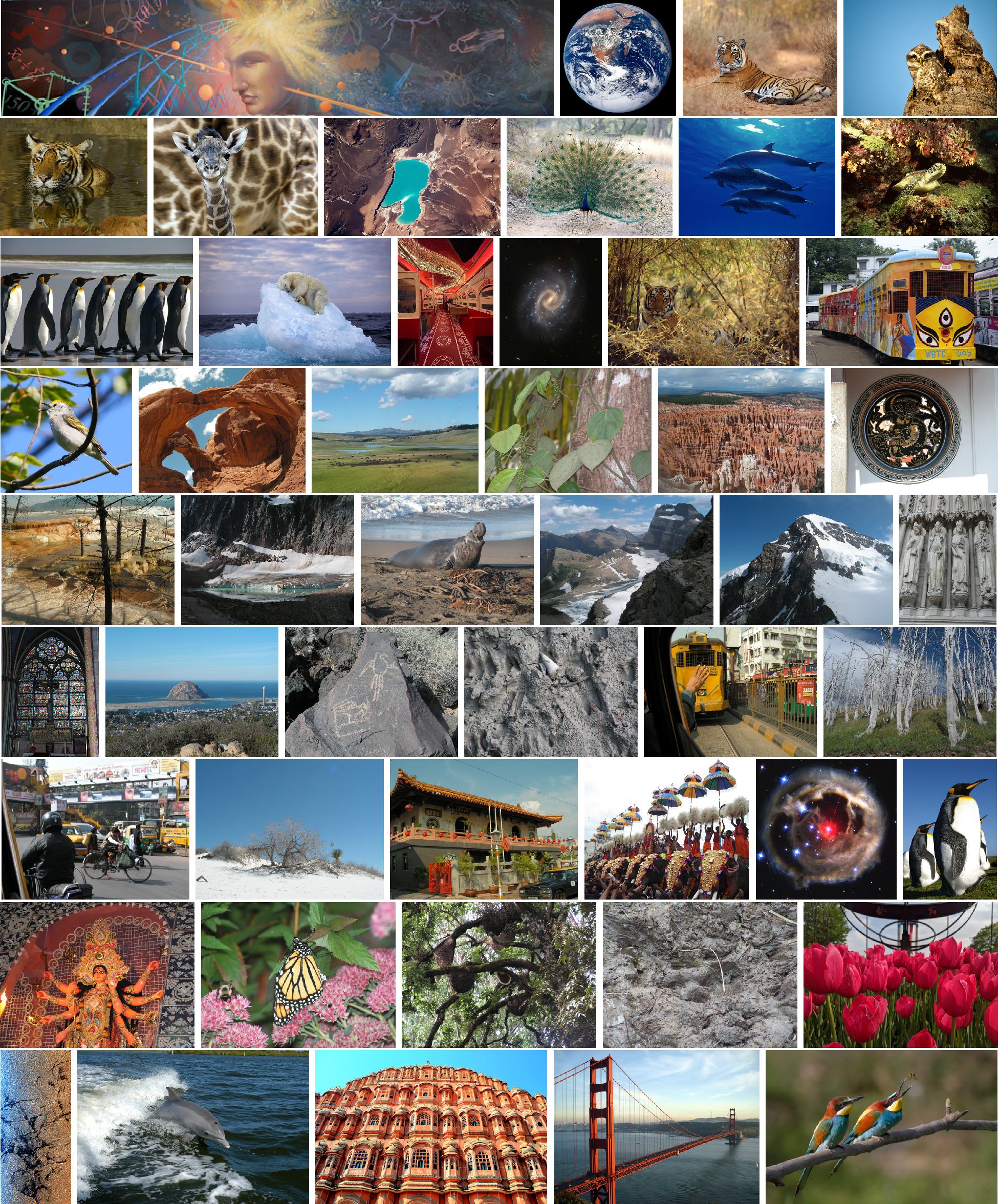}                      
\caption{The smaller-sized images that are Set 1. In nondecreasing order of total number of pixels, they are {\em  snedecormural, BlueMarble, AlertTiger, owlet, Tiger, BabyGiraffe, LagunaVerde, peacock, dolphins, tortoise, MarchingPenguins, IceBed, traminterior, ugc12295, Bandhavgarh, 2023Pujatram, hybridbird, arches, bisons, blackpepper, bryce, ChengHoonTeng, cleopatra, edithcavill, elephantseal, Glacier, Jungfrau, notre-dame-wall, NotreDameAltar, morro, Petroglyphs, sunderbans, tram, trees, ultadanga, white-sands, XiangLinSi, thrissur, ExplodingStar, Penguins, SummerySerenity, munnar, periyar,  TulipsOfPella, bombayhigh, NASAdolphin, hawa, goldengate, Merops}.}
\label{fig:small}
\end{figure}                                                    
Here, we describe the 50 images that comprise the first set displayed in Fig.~\ref{fig:small}.        \begin{itemize}[noitemsep,nolistsep,leftmargin=*]
	\item {\em snedecormural}: This $730{\times}154$ image is a digital picture of the ``Inferences Drawn'' mural at the entrance of Snedecor Hall that is the home of Iowa State University’s Department of Statistics. Installed at the entrance in the main hallway in 2009 as part of the then Art in State Buildings state law, the nearly $25{\times}5$ ft oil-on-canvas mural was drawn by artists Tom Rosborough and Bill Barnes over 18 months and depicts an androgynous human head in the center with over 3,000 symbols, with the ones on the right representing indiscipline and chaos that become thematically simplified and defined as they get  processed by the human head in the center. There are a few versions of this image available online, but we use the one available as part of the VIF package. 
\item {\em BlueMarble}: Iconic $540{\times}540$ view of the Earth on December 7, 1972  as seen by the Apollo 17 crew traveling toward the moon, and marking the first photograph of the south polar ice cap. 
\item {\em AlertTiger}: This $800{\times}600$ image of a resting but alert tiger {\em Panthera Tigris} in Ranthambore National Park (NP) or RNP, India, is from the popular  \url{www.indiamike.com} 
travel forum.
\item {\em IndianSpottedOwlet}: This $800{\times}604$ image is of the common Indian nocturnal bird {\em Athene brama} or spotted owlet, basking in the late evening sunlight, when photographed by Mymoon Moghul.
	\item {\em T17}: This $936{\times}731$ Koshy Koshy image is of a swimming tigress codenamed T17 at RNP.
	\item {\em BabyGiraffe}: This $1000{\times}713$ image is of a baby giraffe born in 2017 at the Houston zoo.
	\item {\em LagunaVerde}: Astronaut Woody Hoburg' $1041{\times}694$ photograph               
of Laguna Verde from the International Space Station as it orbited 264 miles (425 km) over it, July 7, 2023.                  
\item {\em peacock}: This $1024{\times}725$ Koshy Koshy image is of a male peacock in courtship at RNP.
\item {\em dolphins} and {\em tortoise}: $1024{\times}768$ images of Atlantic spotted dolphins and tortoise in  deep waters.
\item {\em MarchingPenguins}: This $1200{\times}797$ image is a medium-resolved adaptation of a larger digital photograph of King penguins {\em Aptenodytes patagonicus patagonicus} in the West Falkland Islands. 
\item {\em IceBed}: Nima Sarikhani's dreamy $1200{\times}800$ photograph, named ``Ice Bed'', of a young polar bear drifting to sleep on a bed carved into an iceberg, first published on February 7, 2024. 
\item {\em traminterior}: This $900{\times}1200$ image, available as part of the VIF package, photographs the interior of a specially decorated tram commemorating both the 150th anniversary of Calcutta Tramways and the recent award of the UNESCO's intangible cultural heritage tag to Kolkata's (formerly, Calcutta) famed Durga Puja festival. The tram exterior is in {\it 2023Pujatram} below.
\item{\em ugc12295}: This $1041{\times}1301$ image is of the spiral galaxy UGC 12295 from the Hubble telescope.
\item {\em Bandhavgarh}: This $1484{\times}991$ image, in the VIF package, is  courtesy of Staffan Widstrand of the World Wildlife Foundation, and shows a Royal Bengal tiger at Bandhavgarh NP, India. 
\item {\em 2023Pujatram}: Also in the VIF package, this $1600{\times}1066$ image accompanies {\em traminterior}.
\item {\em hybridbird}: This $1363{\times}1292$ image, from the Cornell Lab of Ornithology, is of a rare triple-hybrid bird (Golden-winged Warbler, Blue-winged Warbler and Chestnut-sided Warbler)~\citep{toewsetal18}. The digital picture is credited to Lowell Burkett who also reported the finding of the bird.

	The next 20  $1600{\times}1200$ or $1200{\times}1600$ digital photographs are courtesy the author (RM).
\item {\em arches}: This 2007 photograph is of a double arch in Arches NP, USA. 
\item {\em bisons}: This 2008 photograph is of a herd of bison  in Hayden Valley,  Yellowstone NP (YNP), USA.
\item {\em blackpepper}: Black pepper {\em pepper nigrum} plant in village along the backwaters of Kerala, India.
\item {\em bryce}:  This 2007 photograph is of the Hoodoos Amphitheater at Bryce NP, USA.
\item {\em ChengHoonTeng}: 2005 photograph of a wall from Melaka's 1673-built Cheng Hoon Teng Temple that is Malaysia's oldest Chinese temple practising Buddhism, Confucianism and Taoism. 
\item {\em cleopatra}: This 2008 photograph is of Cleopatra terrace in Mammoth Hot Springs in  YNP. 
\item {\em edith-cavill}: This 2010 photograph is of Angel Glacier and Cavell Pond, in Jasper NP, Canada. 
\item {\em elephantseal}: A bloodied elephant seal in 2008 at Piedras Blancas Rookery, California.
\item {\em Glacier}: Grinnell Glacier in 2008, from atop the Garden Wall in Glacier NP, USA. 
\item {\em Jungfrau}: This view in 2010 is from Jungfraujoch, Europe's highest accessible railway station.
\item {\em notre-dame-wall} and {\em NotreDameAltar}: Kings of the Old Testament on the west facade, and stained glass window and altar in ambulatory chapel of Notre Dame Cathedral, Paris, July 2010.
\item {\em morro}: This 2009 view of the {\em El Morro} Rock is from Black Hill in Morro Bay, California.
\item {\em Petroglyphs}: Two petroglyphs at Boca Negra Canyon, Petroglyph National Monument, 2008.
\item {\em sunderbans}: Tiger pug marks in the Sundarbans Tiger Reserve, West Bengal, India, 2010.
\item {\em tram}: A taxi driver waves at a tram driver to allow his vehicle  to cut in, Kolkata, 2010.
\item {\em trees}: Dead, bleached trees at Mt. Washburn after a native forest fire, YNP, 2008.
\item {\em ultadanga}: Multimodal traffic chaos at the Ultadanga-VIP Road interstection, Kolkata, 2010.
\item {\em white-sands}: A Rio Grande cottonwood tree at White Sands NP, USA in 2008.
\item {\em XiangLinSi}: The Xiang Lin Si Temple as an example of a typical Fujian temple in Melaka, December 2005. 
\item {\em thrissur}: A 1619${\times}$1384 Rajesh Kakkanatt photograph of the Thrissurpooram festival in India.
\item {\em ExplodingStar}:  This $1651{\times}1651$ image from the Isaac Newton Group of telescopes is of {\em V838 Monocerotis} which exploded in January 2002 to became the brightest star in our galaxy.
\item {\em Penguins}: King Penguins in Falkland Islands are in this $1361{\times}2048$ photograph by Ben Tubby.

The next four $2560{\times}1920$ digital photographs are from the author's collection.
\item {\em durga}: Decorations, many made of loofahs or pith, at Kolkata's 2012 Durga Puja festival.
\item {\em SummerySerenity}: Monarch and bumblebee at work on a summer day in Ames, Iowa, 2011.
\item {\em munnar}: Beehives hanging from a tree in Munnar, Tamil Nadu, India, 2011.
\item {\em periyar}: Tiger pug marks in the mud at Periyar NP, Kerala, India, 2011.
\item {\em TulipsOfPella}: A $2592{\times}1944$ Subrata Pal (SP) photograph of tulips in Pella, Iowa, 2023.
\item {\em bombayhigh}: NASA's $1644{\times}3604$ radar image of an offshore drilling field in the Arabian Sea. 
\item {\em NASAdolphin}: A $3000{\times}1995$ photograph of a bottlenose dolphin {\em Tursiops truncatus} surfing the wake of a research boat on the Banana River - near NASA's Kennedy Space Center.
\item {\em hawa}: Manudavb's $3456{\times}2304$ photograph of the facade of the Hawa Mahal in Jaipur, India.
\item {\em goldengate}: Brock Brannen's $3264{\times}2448$   photograph of the Golden Gate Bridge, California. 
\item {\em Merops}: Pierre Dalous' $3570{\times}2378$ photograph of two European bee-eaters {\em Merops apiaster}, where the female (in front) awaits the offering that the male will make upon catching the bee. 
\end{itemize}
\subsection{The second set of images}
\begin{figure}
  \centering
  \includegraphics[width=\textwidth]{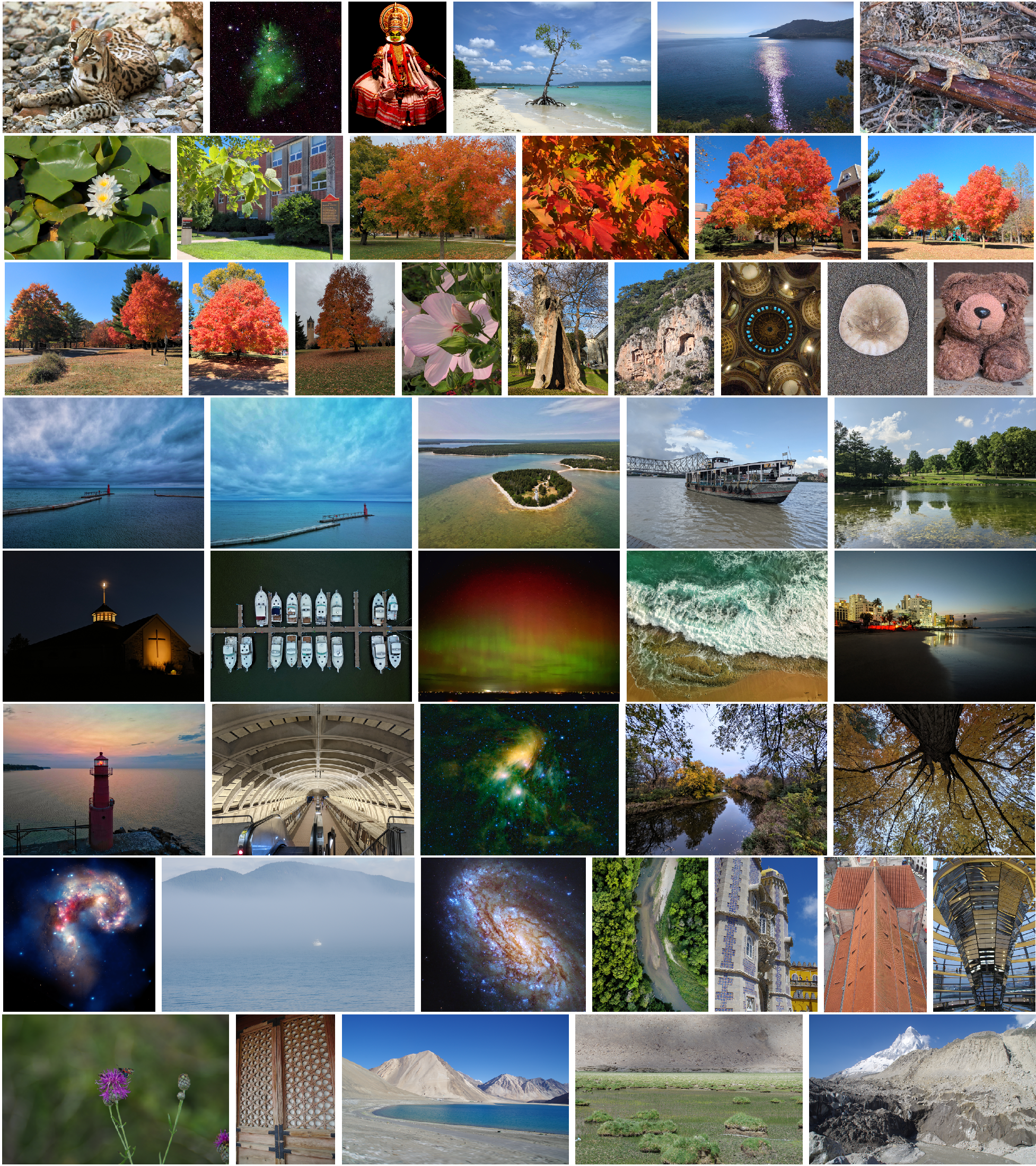}
  \caption{The moderate-sized images that form Set~2: {\em ocelot, ngc2264, kathakali, MangroveTree, MediterraneanRipples, CoastRangeFenceLizzard, WaterLilies, snedecorhall, FallRiot, TurningLeaves, FallBeforeBlack, FallFrenzy, FallFiesta, AristocraticAutumn, StagesOfFall, StagesOfBloom, IndomitableTopkapiTree, KaunosTombsOfTheKings, CirclesInStPauls, SandDollar, LittleSmokeyBear, Azure, Blue, CannaIsland, HooghlyFerry, LaVerne, CrossedCrescent, Moored, NorthernLights, PuertoRico, SanJuan, TheSentinel, FrayedSymmetry, PIA13121, FallSolitude, Interrupted, PIA13316, FoggySound, ngc4654, SSkunkRiver, sintra, OldPeter, ReichstagDome, Nectar, changgyeonggung, Pangong, Ladakh, gomukh.}}
\label{fig:medium}
\end{figure}
We now describe the 48 images that comprise the second set displayed in Fig.~\ref{fig:medium}.\begin{itemize}[noitemsep,nolistsep,leftmargin=*]
\item {\em ocelot}: A $3658{\times}2400$  US Fish and Wildlife Service photograph of an ocelot {\em Leopardus pardalis}.  
\item {\em ngc2264}: 
	This $3000{\times}3000$ image of NGC2264 or the ``Christmas Tree Cluster'' is from  NASA's Chandra X-ray observatory, released on December 19, 2023. 
\item {\em kathakali}: This $2658{\times}3547$ photograph is of a male {\em Kathakali} dancer in India. 
\item {\em MangroveTree}:  Vyacheslav Argenberg's 2009 $4032{\times}2688$ photograph of a solitary mangrove tree on the beach of Havelock Island in the Andaman Sea, Andaman and Nicobar Islands, India. 
\item {\em MediterraneanRipples}: RM's cropped $4032{\times}2079$ photograph of sunlight rippling on the waters of the Mediterranean, off the Dat\c ca peninsula of Turkey (TR), December 2024.

	The next 16 and 12 $3024{\times}4032$ or $4032{\times}3024$ photographs are by RM and SP, in that order.
\item {\em CoastRangeFenceLizzard}: A coast range fence lizzard just off the beach in Morro Bay, USA, 2023.
\item {\em WaterLilies}: Waterlilies among the greens, Kirkland, Washington, August 2024.
\item {\em snedecorhall}: The front of Snedecor Hall, home of ISU's Department of Statistics, August 2023.
\item {\em FallRiot}: A riot of color across the road from ISU's Snedecor Hall, October 2023.
\item {\em TurningLeaves}: Maple leaves changing color in front of Snedecor Hall, October 2023.
\item {\em FallBeforeBlack}: Fall color in front of Black Hall at ISU campus, October 2023.
\item {\em FallFrenzy}, {\em FallFiesta} and {\em AristocraticAutumn}:  Fall in Ames, Iowa, October 2024.
\item {\em StagesOfFall}: Fall  2024 on and around a tree backgrounded by ISU's 125 year-old Campanile.
\item {\em StagesOfBloom}: Rose hibisci in different stages of bloom in a Ames, Iowa front yard, July 2023.
\item {\em IndomitableTopkapiTree}: Tree with hollowed trunk in Topkapi Palace, Istanbul, December 2024. 
\item {\em KaunosTombsOfTheKings}: Tombs of the Kings, from R. Dalyan in Kaunos, TR, December 2024.
\item {\em CirclesInStPauls}: Inside St. Paul's Cathedral, London, United Kingdom, December 2024.
\item {\em SandDollar}: A sand dollar swept on to the beach by the Pacific in Morro Bay, California, 2023.
\item {\em LittleSmokeyBear}: A stuffed grizzly bear from a Many Glacier gift store at Glacier NP, 2008.
\item {\em Azure} and {\em Blue}: Two aerial views in 2023 of  Algoma Pierhead Lighthouse, Wisconsin (WI).
\item {\em CanaIsland}: Drone-captured aerial view of Canna Island, WI, with lighthouse built in 1869.
\item {\em HooghlyFerry}: A Kolkata-Howrah ferry boat, with the iconic Howrah bridge in the background.
\item {\em LaVerne}: Drone photograph of Iowa State University's iconic Lake LaVerne, Ames, IA.
\item {\em CrossedCrescent}: Cross on a church's steeple in front of a waxing gibbous moon, Ames, Iowa.
\item {\em Moored}: Aerial view, by drone, of boats moored in the early morning in Canna Island, WI.
\item {\em NorthernLights}: The northern lights seen in this slow-release photograph in Northern Iowa.
\item {\em Roaring}: Drone-captured aerial photograph of a beach in San Juan, Puerto Rico.
\item {\em LightandDusk}: The beaches of the Atlantic at San Juan, Puerto Rico.
\item {\em TheSentinel}: A drone captures Algoma Pierhead Lighthouse, WI with sunrise through the lens.
\item {\em FrayedSymmetry}: View of platforms from inside Metro station, Washington DC.
\item {\em PA13121}: This $4007{\times}3061$ image is of the famous Pleiades cluster of stars as seen through NASA's Wide-field Infrared Survey Explorer (WISE), released on July 16, 2010.
\item {\em FallSolitude} and {\em Interrupted}: SP's $4080{\times}3072$ photographs of Fall 2023 in Ames, Iowa, as captured by a drone of South Skunk River, and separately and up a solitary tree.

\item {\em zvPIA13316}:  A $3545{\times}3600$ composite image from the Chandra X-ray Observatory (blue), Hubble Space Telescope (gold and brown), and  Spitzer Space Telescope (red) of the Antennae galaxies, located about 62 million light years away from earth, and formed by intergalactic collision. 
\item {\em FoggySound}: A $4928{\times}3014$ cropped RM photograph of Puget Sound in the fog, August 2015.
\item {\em ngc4654}: NASA's $4096{\times}3843$ image of the intermediate spiral galaxy, NGC 4654, in the constellation Virgo, about 55 million light years away from earth.
\item {\em SSkunkRiver}: A $3464{\times}4618$ aerial SP photograph of South Skunk River in Ames, IA, 2023.

	The next eight $3264{\times}4928$ or $4928{\times}3264$ photographs are from RM.
\item {\em sintra}: A view of the Pena Palace in Sintra, Portugal, June, 2019.
\item {\em OldPeter} and {\em ReichstagDome}: The clay orange pantile roof of Peterskirche in Old Town Munich, and interior view of the dome in Berlin's Reichstag building, that was the historical seat of Germany's parliament since the Holy Roman Empire but was damaged during Word War II and only fully restored and reconstructed after the reunification of Germany December 2023.
\item {\em Nectar}:  Bees in search of nectar in Varenna, Lake Como, Italy, July 2013.
\item {\em changgyeonggung}:  Wooden door in Changgyeonggung Palace, Seoul, Korea, July 2024.
\item {\em Pangong} and  {\em Ladakh}: Pangong Tso (Lake), and green vegetation amid the arid cold desert en route from Leh, Ladakh to Pangong Tso, Ladakh, India, July 2014.
\item {\em gomukh}:  Gomukh glacier with Mt Sudarshan in the background, Uttarakhand, India, June 2017.
\end{itemize}
\subsection{The third set of images}
\begin{figure}
  \centering
  \includegraphics[width=\textwidth]{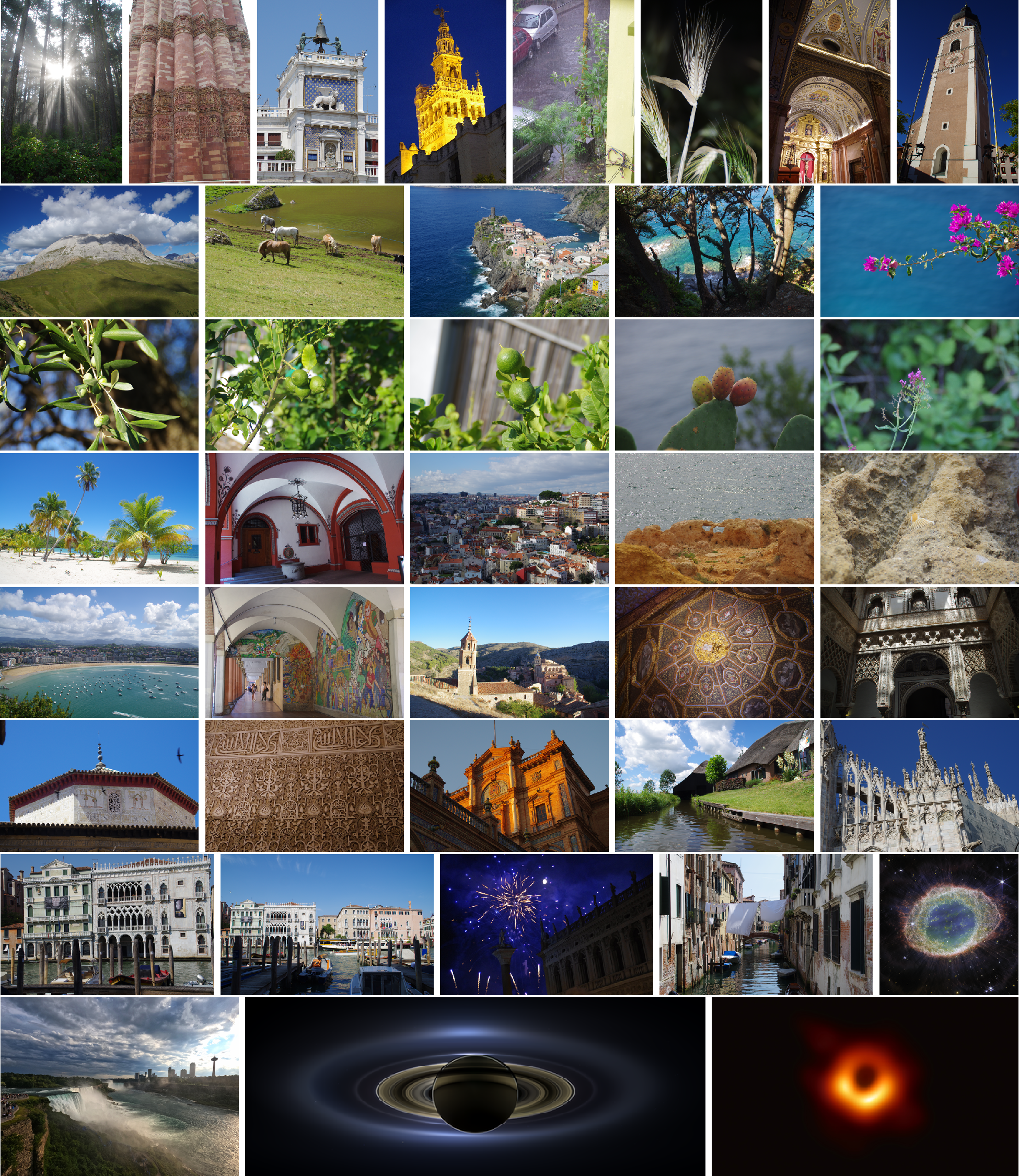}
  \caption{The images in Set 3: {\em FilteredRays, qutb, SanMarcoClockTower, GiraldaNight, Monsoon, GrainsOfTurtuk, macarena, CampanileOfKastelruth, dolomiti, cavalli, vernazza, monterosso-trees, vibrant-colors, olives, limes, 2limes, cactus-nochtli, valeriana, Vieques, Rathaus, lisbon, Algarve, fossil, flotilla, bologna, albarracin, sintratower, alcazar-shadows, alcazar-tower, alhambra-wall, plaza-espana, Giethoorn, milano-duomo, Ca' d'Oro, vaporetti,  Venice, venice-life, RingNebula, Niagara, PIA17172, blackhole}.}
\label{fig:large}
\end{figure}
The third set in Fig.~\ref{fig:large} consists of 44 images, with the first 38 of them  being $3264{\times}4928$ or $4928{\times}3264$ photographs from the author's collection. We describe them briefly.
\begin{itemize}[noitemsep,nolistsep,leftmargin=*]
\item {\em FilteredRays}: Sunlight filters through trees in Anacortes' Washington State Park, August 2015.
\item {\em qutb}: Engravings on the fluted base of the Qutb Minar, New Delhi, India, July 2014.
\item {\em SanMarcoClockTower}: 1 pm on the clock tower in San Marco Square, Venice, Italy, July 2013.
\item {\em GiraldaNight}:  Bell tower of the Seville Cathedral lit up at night, Seville, Spain, June 2019.
\item {\em Monsoon}: From upstairs window in Ultadanga's VIII-M housing complex, Kolkata, July 2013.
\item {\em GrainsOfTurtuk}: Barley stalks cultivated along the Shyok River in Turtuk, India, July 2014.
\item {\em macarena}:  View of the interior of the Basilica de la Macarena in Seville, Spain, June 2019.
\item {\em CampanileOfKastelruth}: The campanile in early morning light, Castelrotto, Italy, August 2022.
\item {\em dolomiti}: Sassolungo range behind the Seiser Alm plateau in the Dolomites, Italy, August 2022.
\item {\em cavalli}: Horses in the Alpine meadows of the Seiser Alm in the Dolomites, Italy, August 2022.
\item {\em vernazza}: Overlooking Vernazza and the Ligurian sea in the Cinque Terre, Italy, August 2022.
\item {\em monterosso-trees}: Trees along the hiking trail from Vernazza to Monterosso, Italy, August 2022.
\item {\em vibrant-colors}: Bougainvillea overlooking the Ligurian sea in Cinque Terre, Italy, August 2022.
\item {\em olives}, {\em limes}, {\em 2limes} and {\em cactus-nochtli}: Fruits in Cinque Terre, Italy, August 2022.
\item {\em valerian}: Red valerian and in search of nectar in the Como region, Italy, July 2013.
\item {\em Vieques}: The idyllic beaches of Vieques, Puerto Rico, May 2015.
\item {\em Rathaus}: Rathaus (town hall) of Basel, Switzerland, June 2019.
\item {\em lisbon}: The city of Lisbon, likely from the Castelo de Sao Jorge, Portugal, June 2019.
\item {\em Algarve} and {\em fossil}: Rock formations and fossil, in Algar Seco, Algarve,  Portugal, June 2019.
\item {\em flotilla}: Boats in the San Sebastian Bay, San Sebastian/Donostia, Spain, August 2016.
\item {\em bologna}: Street art and students in a corridor of University of Bologna, Italy, August 2022.
\item {\em albarracin}: The medieval village of Albarracin in the Teruel Region of Spain, August 2016.
\item {\em sintratower}: Ceiling of the Heraldry room inside Sintra National Palace, Portugal, June 2019.
\item {\em alcazar-shadows} and {\em alcazar-tower}: Inside, and looking up from, the Patio de las Donacellas, or ``Courtyard of the maidens'' in the Real Alc\'azar de Seville, in Seville, Spain, June 2019.
\item {\em alhambra-wall}: Arabic relief in the Nasrid palaces, Alhambra de Granada, Spain, June 2019.
\item {\em plaza-espana}: Facade of a baroque palace in the Plaza de Espana in Seville, Spain, June 2019.
\item {\em Giethoorn}: View of village of Giethoorn or ``Venice of the North'' in Holland, July 2019.
\item {\em milano-duomo}: Statues atop the Duomo or the cathedral church of Milan, Italy, August 2022.
\item {\em Ca' d'Oro}, {\em vaporetto}, {\em Venice}, and {\em venice-life}: 
	Scenes from Venice: the Ca' d'Oro and Palazzo Giusti on the Grand Canal, the Ca' d'Oro vaporetto stop, fireworks at San Marco Square on {\em Festa del Redentore}, and clothes drying acrosss a canal, July 2013.
\item {\em RingNebula}: This $4097{\times}4156$ 
	image, from the near-infrared camera of NASA’s James Webb Space Telescope, shows the Ring Nebula in unprecedented detail. Formed by a star throwing off its outer layers as it runs out of fuel, the Ring Nebula is an archetypal planetary nebula with a filament structure of the inner ring. The nebula has about 20,000 hydrogen-rich dense globules and an inner region of very hot gas.. 
	The image shows the main shell to have a thin ring of enhanced emission from carbon-based molecules known as polycyclic aromatic hydrocarbons.
\item {\em Niagara}: SP's $5184{\times}3888$ wide-view drone photograph of Niagara Falls, Canada, August, 2023.
\item {\em PIA17172}: Captioned ``The Day the Earth smiled,'' this $9000{\times}3500$ image of Saturn released by NASA was reconstructed from Cassini that on July 19, 2013 slipped into Saturn's shadow and turned to image the planet, its inner rings and moons and Earth. 
\item {\em blackhole}: This $7416{\times}4320$ image \citep{akiyamaetal19} is the first ever made of a black hole, revealing a fiery doughnut-shaped object in a galaxy 53 million light-years from Earth. Assembled from data from eight radio telescopes around the world, the image shows light and gas swirling around the lip of a supermassive black hole, a monster of the universe whose existence was theorized by Einstein more than a century ago but confirmed only indirectly over the decades.
\end{itemize}

\section*{Acknowledgments}
This work was initiated as part of a 2016 MS creative component project of Juan P. Rodriguez-Ramirez at Iowa State University: his help in the initial stages of the project is gratefully acknowledged.

%

\bibliography{bibliography}
\bibliographystyle{IEEEtran}

\end{document}